\documentclass{aa}  

\usepackage{graphicx}

\usepackage[varg]{txfonts}
\usepackage{lscape}

\usepackage{natbib}
\bibpunct{(}{)}{;}{a}{}{,}             

\defcitealias{ka10}{KP10}
\defcitealias{kru08}{KM08}

\newcommand{\cmmth}{\mbox{cm$^{-3}$}}
\newcommand{\cmmtw}{\mbox{cm$^{-2}$}}

\newcommand{\kms}{\mbox{km\,s$^{-1}$}}
\newcommand{\msun}{\mbox{$M_{\odot}$}}

\newcommand{\microns}{\mbox{$\mu$m}}
\newcommand{\hii}{H{\scriptsize II}}
\newcommand{\uchii}{UCH{\scriptsize II}}

\begin{document}

\titlerunning{The mass distribution of IRDCs}
\authorrunning{L. G\'omez et al.}

   \title{The mass distribution of clumps within infrared dark clouds. A Large APEX Bolometer Camera study
             \thanks{Based on data acquired with the Atacama Pathfinder Experiment (APEX). APEX is a collaboration between the Max-Planck-Institut f\"ur Radioastronomie, the European Southern Observatory, and the Onsala Space Observatory.}  \fnmsep \thanks{Tables 3 and 4 are only available in electronic form at 
http://www.aanda.org}
}

   \author{
          L. G\'omez\inst{1,2,3}
          \and
          F. Wyrowski\inst{2}
          \and
          F. Schuller\inst{4}
          \and
          K. M. Menten\inst{2}
          \and
          J. Ballesteros-Paredes\inst{5}
          }

   \institute{
             Departamento de Astronom\'\i a, Universidad de Chile, Camino El 
             Observatorio 1515, Las Condes, Santiago, Chile\\
             \email{lgomez@das.uchile.cl}
             \and
             Max-Planck-Institut f\"ur Radioastronomie, Auf dem H\"ugel 69,
             D-53121 Bonn, Germany\\
             \email{[wyrowski, kmenten]@mpifr.de}
             \and
             CSIRO Astronomy and Space Science, P.O. Box 76, Epping NSW 1710, 
             Australia
             \and
             European Southern Observatory, Alonso de C\'ordova 3107, 
             Vitacura, Santiago, Chile\\
             \email{fschulle@apex-telescope.org}
             \and
             Centro de Radioastronom\'\i a y Astrof\'\i sica, Universidad 
             Nacional Aut\'onoma de M\'exico, Apdo. Postal 3--72, 58090 
             Morelia, Michoac\'an, M\'exico\\
             \email{j.ballesteros@crya.unam.mx}
             }

   \date{Received ; accepted }

 
  \abstract
   {}
   {We  present an analysis of the dust continuum emission at 870~$\mu$m
     in order to investigate the mass distribution of clumps within infrared dark 
     clouds (IRDCs).}
   {We map six IRDCs with the Large APEX BOlometer 
CAmera (LABOCA) at APEX, reaching an rms noise level of $\sigma_{\rm rms}$ = 28--44~mJy~beam$^{-1}$. 
The dust continuum emission coming from these IRDCs was decomposed 
by using two automated algorithms, {\it Gaussclumps} and 
{\it Clumpfind}. Moreover, we carried out single-pointing observations of the 
N$_2$H$^+$ (3--2) line toward selected positions  to obtain kinematic 
information.}
   {The mapped IRDCs are located in the range of kinematic distances of
2.7--3.2 kpc.  We identify 510 and 352 sources with {\it Gaussclumps} and 
{\it Clumpfind}, respectively, and estimate masses and other physical 
properties assuming a uniform dust temperature. The mass ranges are
6--2692~\msun~({\it Gaussclumps}) and 7--4254~\msun~({\it Clumpfind}) and the 
ranges in effective radius are $\sim$0.10--0.74~pc ({\it Gaussclumps}) and 
0.16--0.99~pc ({\it Clumpfind}). The mass distribution, independent of the 
decomposition method used, is fitted by a power law,
 $dN/dM \propto M^{\alpha}$, with an index of $-1.60 \pm 0.06$, consistent
with the CO mass distribution and other high-mass star-forming regions. }
   {}  


   \maketitle
%

\section{Introduction}

     Stars form in the densest parts of molecular clouds.
     The mass distribution ($dN/dM \propto M^{\alpha}$) of dense clumps, 
     which is obtained from observations of lines from the CO and CS 
molecules, is a 
     power law with an index 
     of $\alpha \sim -1.7$ \citep[e.g.,][]{la91,bl93,kr98}.
     Continuum observations of high-mass star formation regions (HMSFRs) in 
     the  millimeter regime have also revealed indices of $\alpha \sim -1.7$ 
     \citep[e.g.,][]{mo04,mu07,lo11}.  When studying the mass distributions of 
     lower mass dense cores, some authors have found a steeper value for the 
     slope of
     that distribution \citep[$\alpha \sim -2.3$; e.g., ][]{be11}, 
     resembling the index of the stellar initial mass function (IMF), i.e.,
     $\alpha = -2.35$ \citep[][]{sa55}. Similar values have been claimed 
     for the mass
     distributions of embedded, open, and globular clusters 
     \citep[e.g,][]{el97,la03}.
These similarities might  
     imply that the form of the mass distribution is carried from the
     interstellar matter to stars. However, 
     the origin of the mass distribution is still a matter of debate,
     and some authors have disputed its universality
      \citep[e.g.,][]{Bastian+10,hs10,we10}.

\begin{table*}
\begin{minipage}{\textwidth}
  \caption{Pointing centers of IRDCs observed with LABOCA.}
\label{Tlabocapos}
\centering
\renewcommand{\footnoterule}{}
    \begin{tabular}{l c c c }
\hline
\hline
\noalign{\smallskip}

IRDC name &   \multicolumn{2}{c}{Position\footnote{Units of right ascension are hours, minutes, and seconds and units of declination are degrees, arcminutes, and arcseconds.}} & Map 1$\sigma_{\rm rms}$ noise \\ 
          &  RA (J2000)          &  Dec (J2000)   &  (mJy beam$^{-1}$) \\
\hline
G329.03$-$0.2   &   16:00:35.1   &$-$53:13:06  &   37 \\  
G331.38+0.2     &   16:10:25.4   &$-$51:22:57  &   44 \\  
G335.25$-$0.3   &   16:29:36.7   &$-$48:59:08  &   32 \\  
G337.16$-$0.4   &   16:37:49.4   &$-$47:38:50  &   31  \\ 
G343.48$-$0.4   &   17:01:01.0   &$-$42:48:41  &   28  \\ 
G345.07$-$0.2   &   17:05:21.3   &$-$41:25:16  &   37  \\ 
\hline        
\end{tabular} 
\end{minipage}
\end{table*}

Several studies reveal that some
    clumps within the so-called infrared dark clouds (IRDCs) harbor the earliest
    stages of HMSF 
    \citep[e.g.,][]{ra06,ba10}. Therefore, IRDCs are good candidates for 
    studying
    the mass distribution of their clumps/cores.
    Different studies of the mass distributions of IRDCs based on either
    extinction or continuum maps and different assumptions, e.g., 
    source extraction, mass estimation, etc., find a
    range of indices from $ \alpha \sim -1.7$ to $-2.1$
    \citep[e.g.,][]{ra06,ra09,pe10b,mi12}.

     Here we aim at studying the mass distribution of clumps within six IRDCs 
     in the dust continuum emission at 870~\microns~that are located
     at kinematic distances of 2.7--3.2 kpc.
     In Sect.~\ref{laboca:obs}, we present the IRDC sample, dust continuum 
     emission observations, and molecular line (N$_2$H$^+$ 3--2) observations. Section~\ref{laboca:results}
     shows the resulting dust emission maps and sample spectra, while Sect.~\ref{laboca:ana} 
     presents the source decomposition analysis,
      the mass-radius relationship, and the clump mass distribution.
The results are discussed in Sect.~\ref{laboca:dis} and, finally, we summarize
      our findings in Sect.~\ref{laboca:sum}.

We adopt a nomenclature in which clumps refer to structures with diameters
   of $\sim$1 pc, embedded within a cloud (of several pc), and cores refer to 
   structures within a clump with diameters of $\sim$0.1~pc.

\section{Observations and data reduction}\label{laboca:obs}

\subsection{Continuum data}\label{laboca:obs:con}

      We selected six southern  hemisphere clouds with high contrast,
        i.e., high extinction, from our
        IRDC catalog at 24~$\mu$m, which was created by applying the method from
        \citet{si06} to  {\it Spitzer}/MIPSGAL images \citep{ca09}.
       In Table~\ref{Tlabocapos} we list the IRDCs observed  
         with LABOCA on the APEX 12m telescope
        \citep[][]{gue06}. 
        LABOCA, the Large APEX BOlometer CAmera \citep[][]{si09}, is a
      bolometer array of 295 pixels working at  870~$\mu$m (345~GHz), a
      bandwidth of about 60~GHz, and a field of view of 11\farcm4.
      The LABOCA instrument was developed by the Max-Planck-Institut f\"ur 
      Radioastronomie.

         The observations were performed 
         on 2007 August 25 and 27-28, typically 
         covering an area 
        of $\sim$20\arcmin $\times$20\arcmin~for each IRDC.
        The sky opacity was determined every one to two hours with skydips.
        The focus was optimized on Jupiter. Pointing observations were 
        checked on the source IRAS 16293$-$2422 
\citep[see Appendix A of][]{si09},
        and the telescope pointing was found to be accurate to within 5\arcsec.

 We used the Bolometer array data Analysis  package \citep[BoA;][]{sc12} to reduce the LABOCA data. The data reduction 
        involves flux calibration, flagging bad and noisy pixels, removal of 
        correlated 
        noise, despiking, low-frequency filtering, and first-order baseline 
        removal. These procedures 
        are explained in detail in \citet{si09} and \citet{sc09}. The 
        removal of correlated noise was done on all pixels with the median 
        noise method with five iterations and a relative gain of 0.8.
        The correlated noise for the pixels sharing the same electronics
        subsystem was removed with two iterations and a relative gain of 0.8.
        The same numbers of iterations and gain were applied to groups of pixels
        connected to the same read-out cable.

        Each map was built using natural weighting, where each data point has
        a weight 1/$\sigma_i^2$ and where $\sigma_i$ is the standard deviation 
        of each pixel.
        The resulting map after performing the whole process was used as a 
        model for the next iteration. A total of 20 iterations were performed. 
        As shown in \citet{be11}, this iterative process 
        helps to recover flux at each iteration and to recover
        extended emission. 
        Still, it is important to point out that the 
         spatial filtering due to the correlated noise removal
        can reduce the sizes of extended structures. \citet{be11} also find that
        the size for their (input artificial) circular 
        sources with sizes from 200\arcsec~to 440\arcsec are reduced by
        15\%~to 50\%, while for elliptical sources with aspect ratio of 2.5 and
        minor axis (FWHM) varying from 19\farcs2 to 222\arcsec, the sizes are
        reduced by 20--25\%.

The final flux calibration is
        accurate to $\sim$20\%.
        The data were projected on maps with a pixel size of one third
 of the beam 
        FWHM and the map in the last iteration
        was smoothed with a Gaussian kernel of 10\arcsec, providing
        a resolution of the final map of 21\farcs6, which corresponds to
        0.30~pc at a distance of 2.9~kpc. The rms  noise 
        level ($\sigma_{\rm rms}$) varies from IRDC to IRDC  between 28 and 
44~mJy~beam$^{-1}$ (see 
        Table~\ref{Tlabocapos}). In
        comparison to the APEX Telescope Large Area Survey of the Galaxy (ATLASGAL), also at 870~$\mu$m \citep{sc09}, our observations
        go deeper by about a factor of 2, which as we  see in the next 
        sections, helps probe  a wider mass range.

\subsection{Molecular line data}\label{laboca:obs:lin}

        In addition to the continuum data, we performed N$_2$H$^+$ (3--2) line
        observations with the double-sideband heterodyne
      receiver APEX-2A  \citep{ri06} toward selected positions in order to
      obtain kinematic distances. In total, 18 positions were observed within
      the six IRDCs, typically with two to five pointings per cloud.
These IR-quiet and IR-loud targets were chosen by eye from 
      {\it Spitzer}/GLIMPSE images at 8 $\mu$m. 
      Table~\ref{Tn2hp:laboca} lists the observed positions and includes 
      whether the target
      is dark at 8 $\mu$m and at 24 $\mu$m.

Observations were carried out with the APEX telescope
      on 2007 October 28-29. The fast Fourier transform spectrometer 
\citep[FFTS;][]{kl06} was used as a backend for these 
      observations. The pointing was checked on the core 
      G327.3-0.6 and the source 18592+0109 every hour. System temperature 
      varied       from 201 to 243 K.
N$_2$H$^+$ (3--2) line parameters\footnote{
Rest frequencies and upper level energies from the CDMS as of February 2011.}\fnmsep \footnote{Main-beam and forward efficiencies are from http://www.apex-telescope.org/telescope/efficiency/index.php} are $\nu$ =  279511.7348 MHz,
      $E_u/k$ = 26.83~K, HPBW = 22\arcsec, $B_{\rm eff}$ = 0.73,
       $F_{\rm eff}$ = 0.95, and $\delta {\rm v}_{\rm res}$ = 0.52~\kms.

 The data was reduced with the CLASS program from the GILDAS package\footnote{http://www.iram.fr/IRAMFR/GILDAS}. 
     We summed up individual scans and fitted and subtracted a polynomial of order 1 or 2 to the 
     baseline of each 
     final spectrum. For several spectra that were affected by standing waves 
     in the optics (between the subreflector and the receiver) or by an 
     instability of the receiver itself, we edited and performed a linear
     interpolation on the Fourier transform to remove the sinusoidal pattern.
     The conversion from $T^*_A$ to $T_{\rm MB}$ was done with the efficiencies
     mentioned before.

\section{Observational results}\label{laboca:results}

\begin{figure*}
\centering
\includegraphics[scale=0.53]{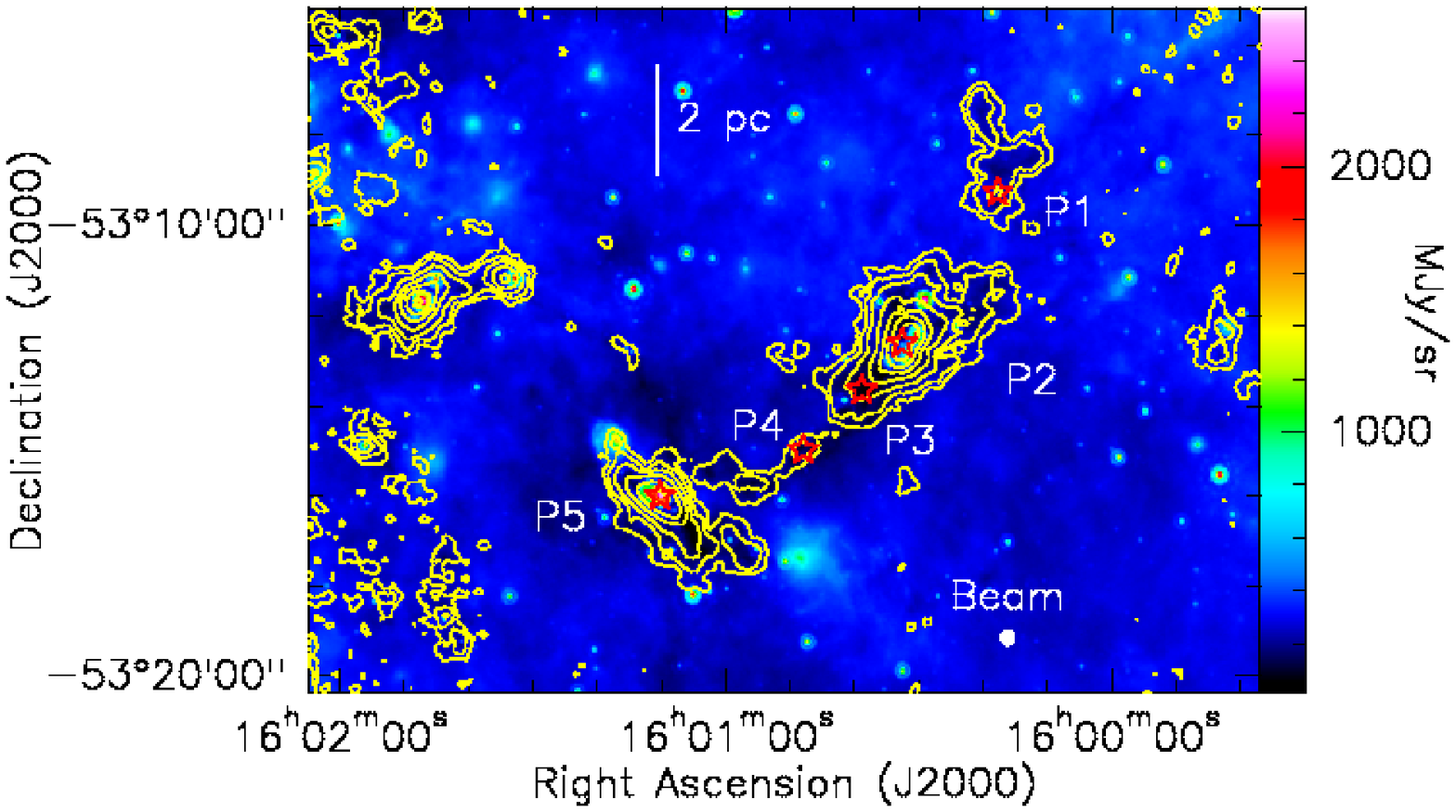}\vspace{0.3cm}
\includegraphics[scale=0.58]{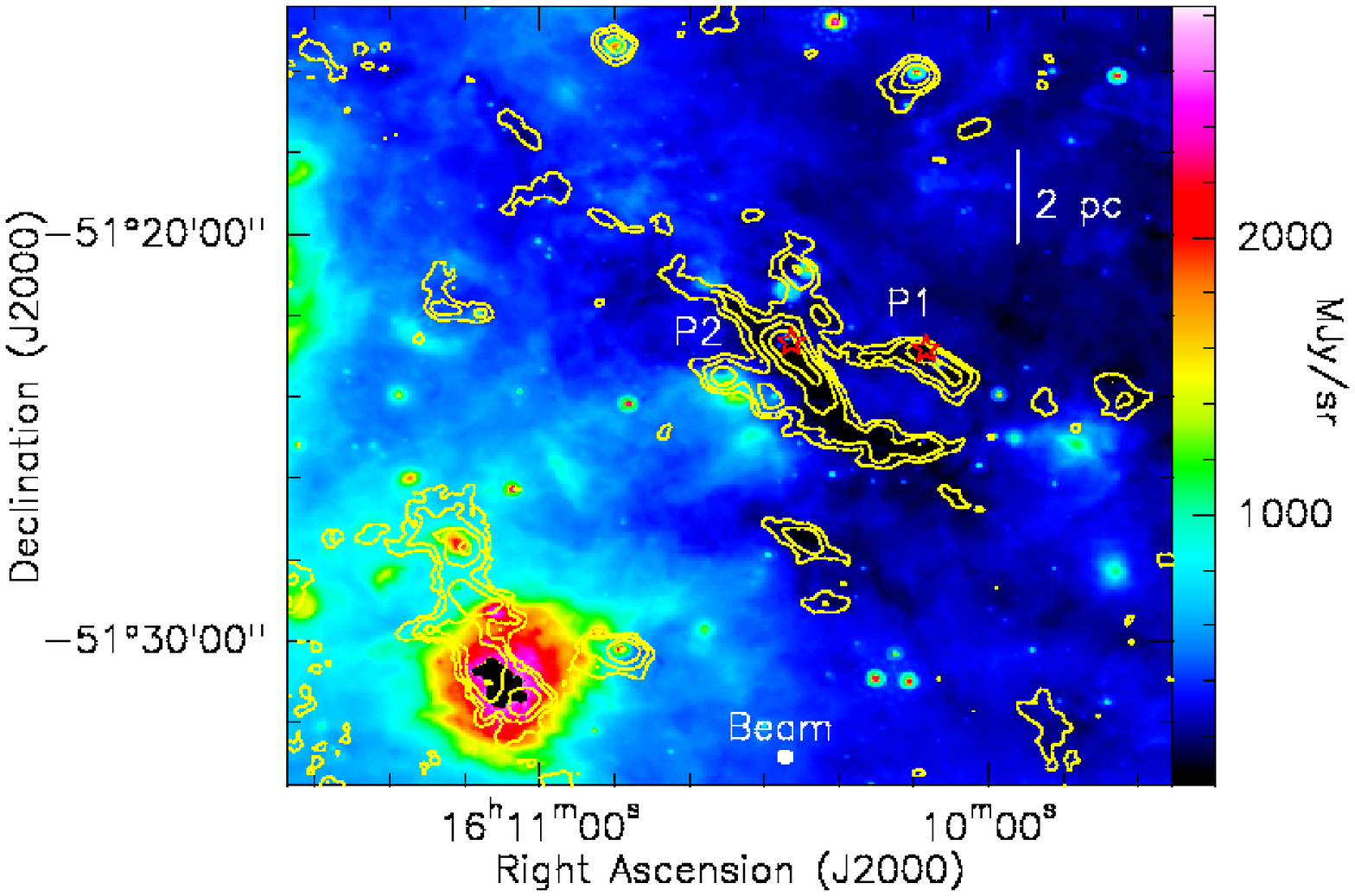}\vspace{0.3cm}
\includegraphics[scale=0.7]{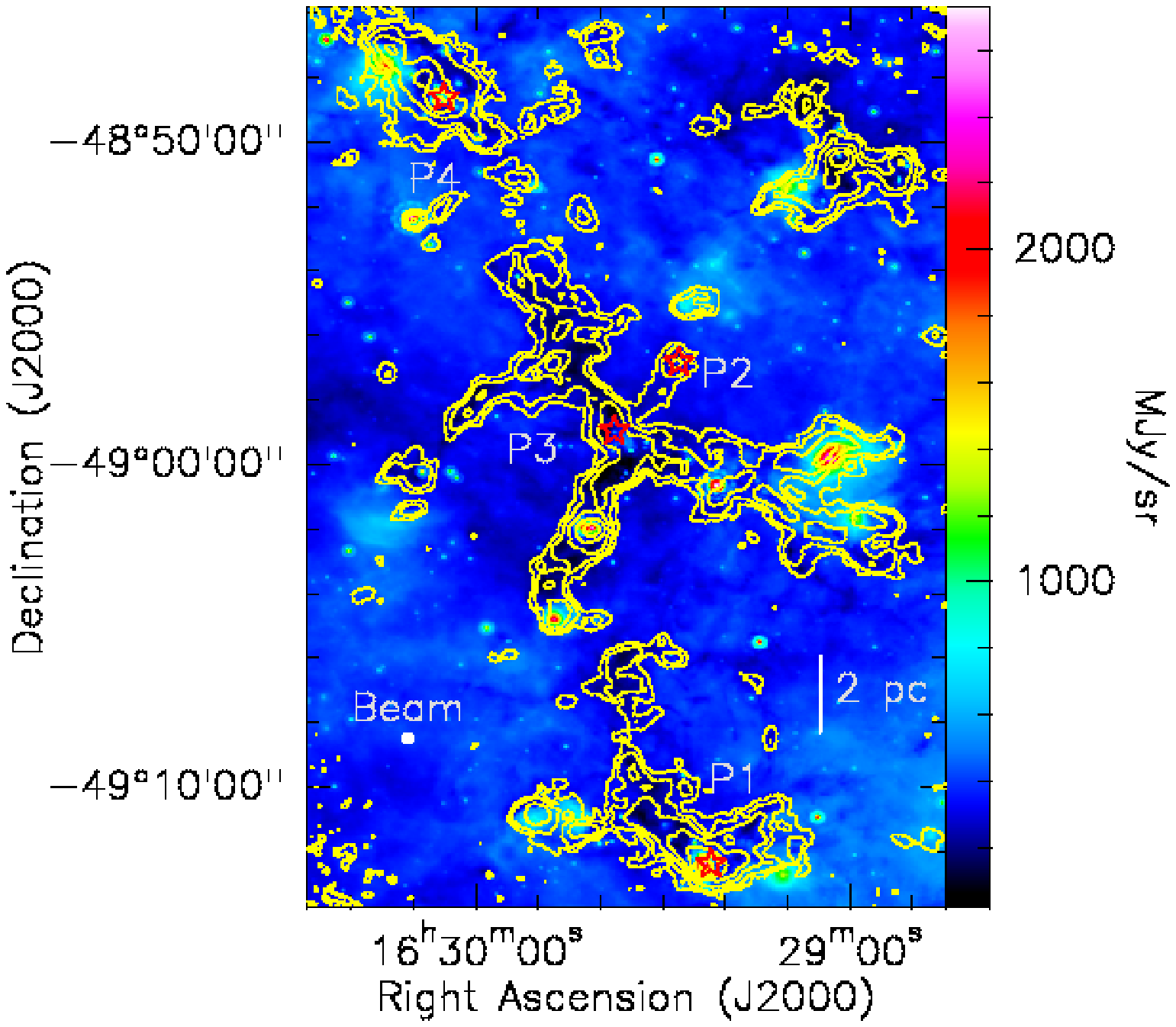}
\caption{
     Contour maps obtained with LABOCA 
     at 870 $\mu$m overlaid on  the {\it Spitzer}/MIPSGAL 24~$\mu$m 
     images for the IRDCs G329 ({\it top}), G331 ({\it middle}), and G335 ({\it bottom}).  The beam {\it HPBW} (21\farcs6) and a 2-pc 
     scale-bar are shown in each panel. Contours for G329 are 3, 6, 12, 24, 48, 96, 192 times 
0.037~Jy~beam$^{-1}$,
     the rms noise of the image. Contours for G331 are 3, 6, 12,
  24 times 0.044 Jy~beam$^{-1}$, the rms noise of the image.
Contours for G335 are 3, 6, 12,
  24, 48, 96, 192 times 0.032 Jy~beam$^{-1}$, the rms noise of the image.
The red stars and the P labels represent the selected positions for molecular
N$_2$H$^+$ (3--2) observations.
}
\label{Firdcslaboca}
\end{figure*}

\begin{figure*}
\centering
\includegraphics[scale=0.6]{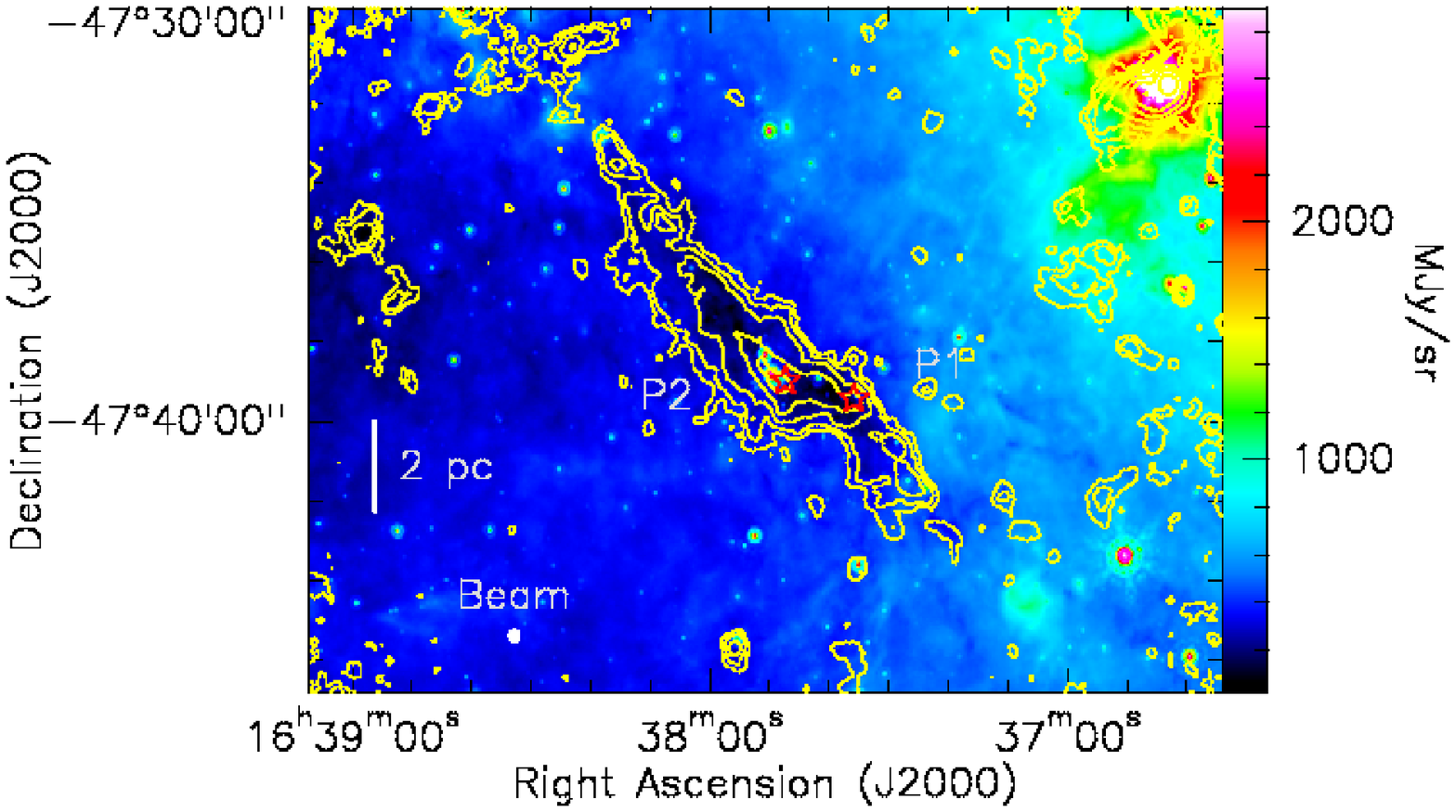}\vspace{0.3cm}
\includegraphics[scale=0.6]{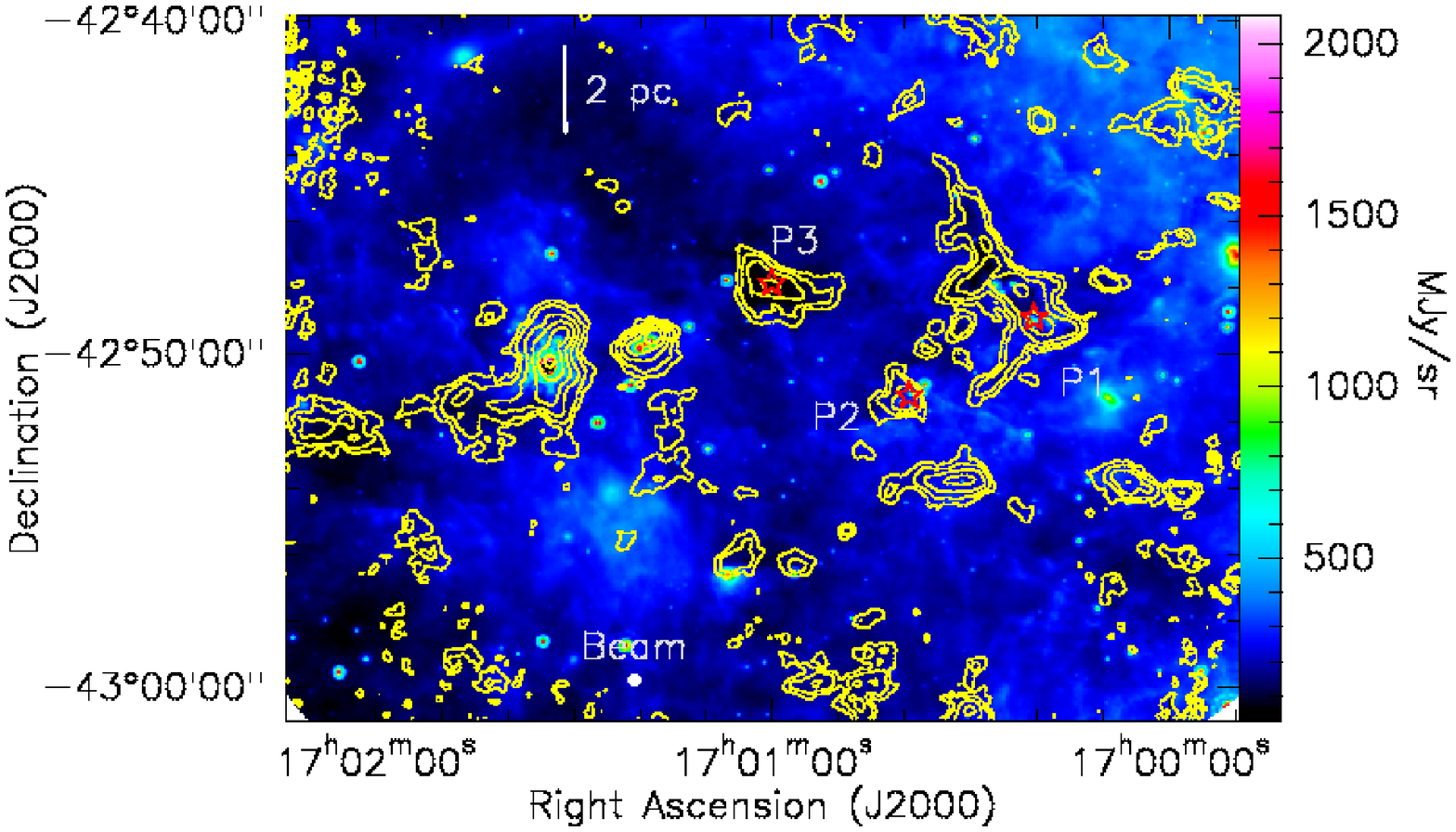}\vspace{0.3cm}
\includegraphics[scale=0.62]{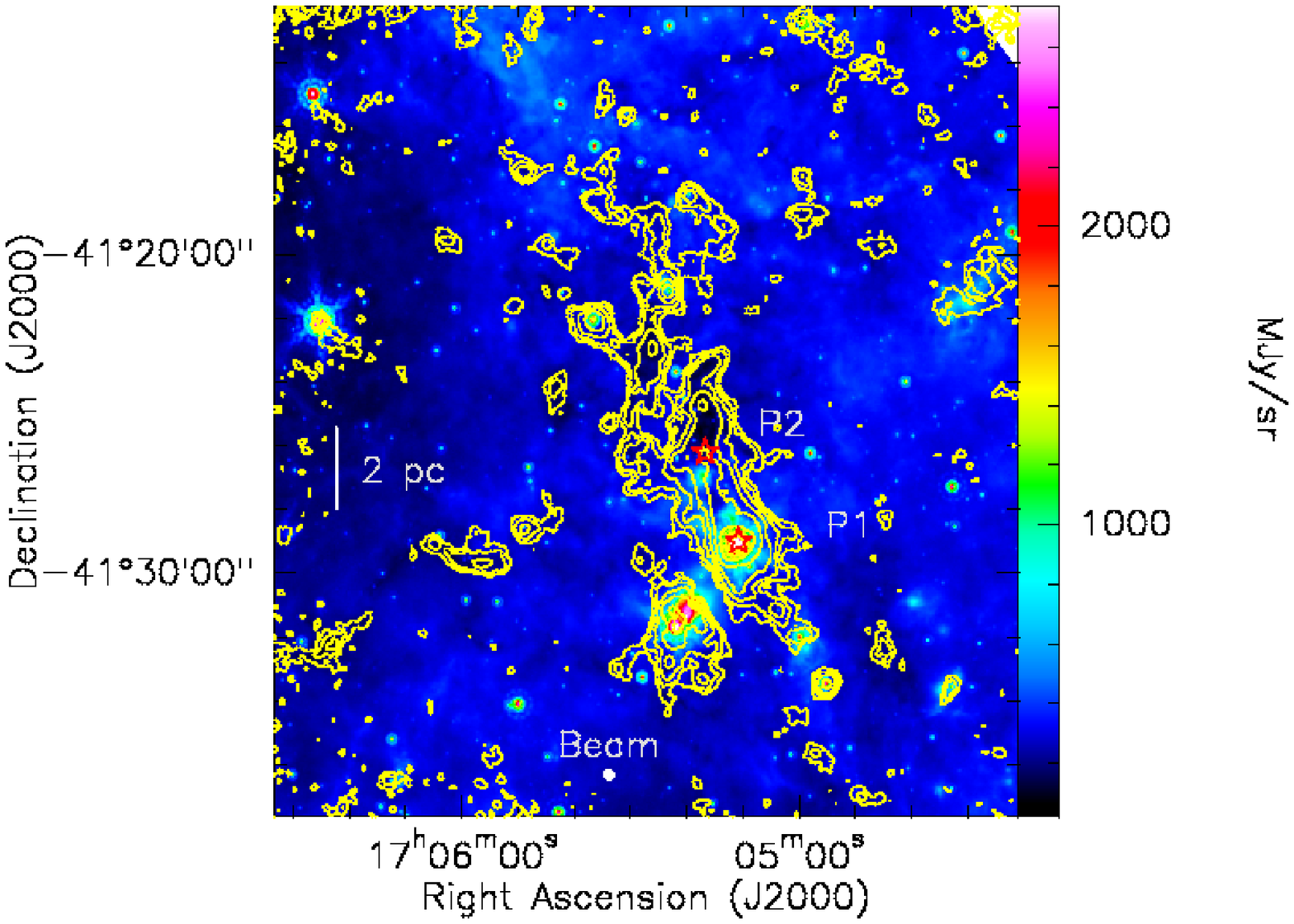}
\caption{
     Same as Fig.~\ref{Firdcslaboca} but for the IRDCs G337 ({\it top}), G343 ({\it middle}), and G345 ({\it bottom}).  
Contours for G337 are 3, 6, 12,
  24, 48, 96, 192 times 0.031 Jy~beam$^{-1}$, the rms noise of the image.
Contours for G343 are 3, 6, 12,
  24, 48, 96 times 0.028 Jy~beam$^{-1}$, the rms noise of the image.
Contours for G345 are 3, 6, 12,
  24, 48, 96, 192, 384 times 0.037 Jy~beam$^{-1}$, the rms noise of the image.
}
\label{Firdcslaboca2}
\end{figure*}

\begin{figure*}
\centering
\includegraphics[scale=0.55]{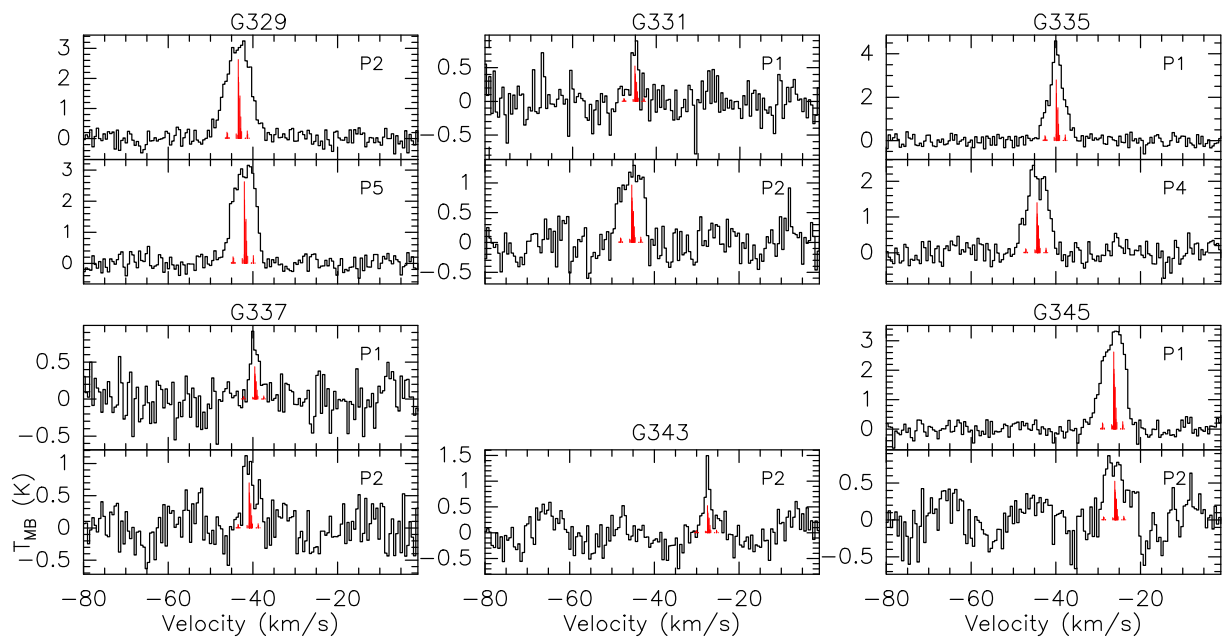}
\caption{
     APEX2A spectra (histograms) of the N$_2$H$^+$ (3--2) line. The red solid 
lines indicate the relative 
      intensities for each of the 29 hyperfine components. P labels refer to the
positions marked in Figs.~\ref{Firdcslaboca}~and~\ref{Firdcslaboca2} with
red stars.
}
\label{Fapex2a-spectra}
\end{figure*}

Figures~\ref{Firdcslaboca} and~\ref{Firdcslaboca2}  show  the final maps 
obtained with LABOCA in contours overlaid on 
     {\it Spitzer}/MIPSGAL 24~$\mu$m images.
The mapped regions show extended, filamentary, and compact dust 
     continuum emission. The prominent features are those associated with bright
     24~$\mu$m sources. Filamentary IR-dark structures in the 24~$\mu$m 
     images can be seen in dust emission at 870~$\mu$m extending for 
     more than 10\arcmin~($\sim$9~pc at a distance of 2.7~kpc). 
     All clouds show several compact sources. Some of these compact sources
     are shrouded in the filamentary emission while others 
     surround the extended structures.

     In general, there is good correlation between the 24~$\mu$m dark structures
     and the emission at 870~$\mu$m, but there are some IR-dark patches that do
     not have a submillimeter emission counterpart. 
     IRDCs G343 and G345 seem to be more fragmented than the others and present
     emission all over the field contrary to, say IRDCs G329 and G331. 
      We see clouds that are connected/linked by an envelope, e.g., IRDCs G337
     and G331; clouds with clear compact sources still embedded
     in the envelope, e.g., IRDCs G329 and G335; and clouds that are more
     dispersed over the mapped regions, e.g., IRDCs G343 and G345.

 Sample spectra of the N$_2$H$^+$ (3--2) line toward 
      positions where emission was detected are shown
in Fig.~\ref{Fapex2a-spectra}.  Gaussian fits results are listed in
Table~\ref{Tn2hp:laboca}; we detected emission in 11 sources and
the average 1$\sigma_{\rm rms}$ noise level for non-detections
     was $\sim$0.24~K. The line widths obtained with the hyperfine structure
fitting procedure of CLASS (``method HFS'')  are narrower than those obtained with Gaussian fits,
up to 50\% in two clumps (G331P2 and G345P1). In these two cases, we obtained 
     very high optical depths ($\sim$13).    
The hyperfine components of N$_2$H$^+$ are also plotted in 
     Fig.~\ref{Fapex2a-spectra}.

\section{Analysis}\label{laboca:ana}

\begin{table*}
\begin{minipage}{\textwidth}
  \caption{N$_2$H$^+$ (3--2) observational parameters, distances, and virial masses.}
\label{Tn2hp:laboca}
\centering
\renewcommand{\footnoterule}{}
    \begin{tabular}{l c c c c c c c  c l r}
\hline
\hline
\noalign{\smallskip}

IRDC & IR-dark\footnote{``Y'' means the clump is 8 $\mu$m- or 24 $\mu$m-dark and ``N'' means the clump is 8 $\mu$m- or 24 $\mu$m-bright.}  &\multicolumn{2}{c}{Position\footnote{Units of right ascension are hours, minutes, and seconds and units of declination are degrees, arcminutes, and arcseconds.}}  &    $T_{\rm MB}$ & $\Delta$V &  v$_{\rm LSR}$ & $D$\footnote{Derived (near) kinematic distances by using the N$_2$H$^+$ (3--2) line (see Sect.~\ref{laboca:ana:kd}).} &  $M_{\rm vir}$\footnote{Virial mass of clumps estimated using two density distributions. The first value corresponds to a constant density distribution while the second value corresponds to a density distribution that varies as $\rho \propto 1/r^2$.} & $N_{\rm cl}$\footnote{Associated {\it Gaussclumps} running number and gas mass (see Sect.~\ref{laboca:decomposition} and Table~\ref{Tgaussclumps}).} &  $M^e$ \\
  clump    &8/24 & RA & Dec  &    &    &   &    &  & &\\
         &       & (J2000)     &(J2000) & (K)  &  (\kms)  & (\kms)  & (kpc)   &   (\msun)  &  & (\msun) \\

\hline 

G329P1  &  N/N  &    16:00:17.3 &   $-$53:09:18 &     (0.25)    &               &                  &       &         &   22   &  209  \\
G329P2  &  Y/N  &    16:00:32.2 &   $-$53:12:39 &    3.20(0.19) &    6.69(0.19) &   $-$43.42(0.08) &  3.0  &  2444/1466   &   1    & 2692  \\
G329P3  &  Y/Y  &    16:00:38.5 &   $-$53:13:41 &      (0.23)   &               &                  &       &         &   12   &  211  \\
G329P4  &  Y/Y  &    16:00:47.7 &   $-$53:15:01 &      (0.25)   &               &                  &       &         &   37   &  56  \\
G329P5  &  N/N  &    16:01:10.1 &   $-$53:16:02 &    3.26(0.32) &    5.83(0.17) &   $-$42.00(0.08) &  2.9  &  1642/985   &   3    &  867 \\
                                                                                                                                 
G331P1  &  Y/Y  &    16:10:08.7 &   $-$51:22:52 &    0.86(0.17) &    1.79(0.50) &   $-$45.10(0.18) &   3.1 &  135/81    &   6    &  168 \\
G331P2  &  N/N  &    16:10:26.9 &   $-$51:22:42 &    1.26(0.25) &    5.60(0.49) &   $-$45.31(0.25) &   3.1 &  1976/1185   &   1    &  553 \\
                                                                                                                                 
G335P1  &  N/N  &    16:29:22.2 &   $-$49:12:24 &    3.88(0.33) &    4.14(0.14) &   $-$39.83(0.05) &   3.0 &  612/367    &   1    &  1002 \\
G335P2  &  Y/N  &    16:29:27.4 &   $-$48:56:51 &      (0.25)   &               &                  &       &         &   32   &   89  \\
G335P3  &  N/N  &    16:29:37.8 &   $-$48:58:58 &      (0.24)   &               &                  &       &         &   13   &  488  \\
G335P4  &  N/N  &    16:30:05.3 &   $-$48:48:37 &    2.29(0.27) &    5.53(0.26) &   $-$44.40(0.12) &  3.2  & 1477/886    &   2    &  664 \\
                                                                                                                                 
G337P1  &  N/N  &    16:37:35.8 &   $-$47:39:27 &    0.78(0.11) &    2.49(0.59) &   $-$39.56(0.24) &  3.0  &  234/140    &   12   &  85  \\
G337P2  &  N/N  &    16:37:47.4 &   $-$47:38:59 &    0.94(0.26) &    3.61(0.72) &   $-$40.85(0.30) &  3.1  &  1095/657   &   3    &  937 \\
                                                                                                                                 
G343P1  &  N/N  &    17:00:20.7 &   $-$42:49:08 &      (0.23)   &               &                  &       &         &   4    &  281  \\
G343P2  &  N/N  &    17:00:39.6 &   $-$42:51:28 &    1.27(0.14) &    1.59(0.50) &   $-$27.42(0.14) &   2.7 &  122/73    &   13   &  104  \\
G343P3  &  Y/Y  &    17:01:00.4 &   $-$42:48:05 &      (0.23)   &               &                  &       &         &   6    &  252  \\
                                                                                                                                 
G345P1  &  N/N  &    17:05:10.5 &   $-$41:29:01 &    3.44(0.33) &    5.85(0.15) &   $-$26.32(0.07) &   2.8 &  1150/690   &   1    &  2630 \\
G345P2  &  Y/Y  &    17:05:16.5 &   $-$41:26:14 &    0.80(0.17) &    5.26(0.76) &   $-$26.04(0.37) &   2.8 &  3137/1882   &   7    &  1520 \\

\hline  
\end{tabular} 
\tablefoot{Parameters are obtained from Gaussian fits. For non-detections, the values given in parenthesis represent the spectra 1$\sigma_{\rm rms}$ noise level.} 
\end{minipage}
\end{table*}

\subsection{Molecular line data}\label{laboca:ana:kd}

      The ``near'' kinematic distances  (see Table~\ref{Tn2hp:laboca}) are estimated based on
      the Galactic rotation curve model by \citet{fi89}, assuming the IAU 
      standard rotation constants of distance to the Sun from 
      the Galactic center as $R_0$ = 8.5~kpc and the Sun's rotation speed
      around the Galactic center as $\Theta_0$ = 220~\kms.
      In the calculations, we use the LSR velocity of N$_2$H$^+$~(3--2).
      In all clouds, except one, we have two detections of N$_2$H$^+$~(3--2). The 
      difference in the LSR velocities  of the clumps will result
      in a distance difference of up to 0.2 kpc, but this difference
      is easily explained by velocity variations within the parental molecular clouds.

\subsection{Source extraction from continuum maps: {\it Gaussclumps} and {\it Clumpfind}}\label{laboca:decomposition}

We use the two most popular algorithms {\it Gaussclumps}
     \citep{st90,kr98} and {\it Clumpfind} \citep{wil94} to extract
     {\it clumps} and derive their physical properties from the  dust emission. 

     {\it Gaussclumps}, a task in the GILDAS\footnote{http://www.iram.fr/IRAMFR/GILDAS} package,
     was originally written to decompose a three-dimensional data cube into 
     Gaussian-shaped sources \citep[see][]{st90} but can also be applied to
     dust continuum maps \citep[e.g.,][]{mo04}. 
     Two adjacent empty planes were
     added to the original two-dimensional maps needed for the algorithm to run
     properly. ``Stiffness'' parameters that control the fitting,
     ensuring that a local {\it clump} is fitted and subtracted, were set to 1
     \citep{kr98}. 
     A peak flux density threshold was set to 5$\sigma_{\rm rms}$. Following \cite{be11},
     the initial guesses for the aperture cutoff, the aperture {\it FWHM}  and 
     the source {\it FWHM} were set to 8, 3, 
     and 1.5 times the angular resolution, respectively.

     The resulting Gaussian sources derived from {\it Gaussclumps} are
     listed in Table~\ref{Tgaussclumps}. 
     Columns are (1) running number
     in the order {\it Gaussclumps} finds the source, (2)--(3) J2000 position, 
     (4) peak intensity, (5) flux density, (6) angular {\it FWHM} along the 
     major and minor axes determined from Gaussian fits before deconvolution, 
     (7) deconvolved {\it FWHM}s (sizes smaller than 
     25.9\arcsec~were set to 25.9\arcsec~to compute the deconvolved sizes, in
     order to account for a fit inaccuracy corresponding to a 5-$\sigma_{\rm rms}$
     detection in peak intensity) and position angle, 
     (8) deconvolved effective radius, $R_{\rm eff}$, (9)--(12) total mass 
     derived from the 
     Gaussian fit, beam-averaged H$_2$ column density, $N_{\rm H_2}$, and 
     volume density, $n_{\rm H_2}$ (see Sect.~\ref{estimates} ).
     Sources lying on noisy edges were discarded from further analysis.
     IRDCs G329, G331, G335, G337, G343, and G345 were 
     decomposed
     into 75, 41, 123, 65, 83, and 123 Gaussian sources, respectively, in 
a total of     510 sources.

\onllongtab{
\begin{longtab}
\small
\begin{longtable}{r c r r r c c r r r c c }
  \caption{Sources found with {\it Gaussclumps} and derived physical parameters.}
\label{Tgaussclumps}\\
\hline
\hline
\noalign{\smallskip}

$N_{\rm cl}$   &  \multicolumn{2}{c}{Position} &   $I_{\rm peak}$  & $S$ &  $\theta_{\rm maj} \times \theta_{\rm min}$ &  {\it FWHM}  & PA  & $R_{\rm eff}$  & $M$  & $N_{\rm H_2}$ & $n_{\rm H_2}$\\
   & R.A.(J2000)   &  Decl.(J2000)   & (Jy/beam) & (Jy) & (\arcsec $\times$ \arcsec) & (pc $\times$ pc) & (\degr) & (pc)  & (\msun) & ($\times 10^{22}~\cmmtw$) & ($\times 10^4~\cmmth$)  \\

\endfirsthead
\noalign{\smallskip}
\caption{continued.}\\
\hline\hline
\noalign{\smallskip}

$N_{\rm cl}$   &  \multicolumn{2}{c}{Position} &   $I_{\rm peak}$  & $S$ &  $\theta_{\rm maj} \times \theta_{\rm min}$ &  {\it FWHM}  & PA  & $R_{\rm eff}$  & $M$  & $N_{\rm H_2}$ & $n_{\rm H_2}$\\
   & R.A.(J2000)   &  Decl.(J2000)   & (Jy/beam) & (Jy) & (\arcsec $\times$ \arcsec) & (pc $\times$ pc) & (\degr) & (pc)  & (\msun) & ($\times 10^{22}~\cmmtw$) & ($\times 10^4~\cmmth$)  \\

\hline
\endhead
\hline

\noalign{\smallskip}

\multicolumn{12}{c}{\bf IRDC G329} \\


1 & 16:00:31.7 & $-$53:12:36 &           12.335  &   52.636  &   60.0 $\times$   33.2  &  0.79 $\times$  0.35  &   $-$28  &  0.26  &  2692  &   27.0  &   50.8  \\
2 & 16:01:47.7 & $-$53:11:44 &            8.848  &   18.485  &   35.2 $\times$   27.7  &  0.39 $\times$  0.24  &   $-$50  &  0.15  &   945  &   19.4  &   89.3  \\
3 & 16:01:10.1 & $-$53:16:04 &            5.034  &   16.961  &   39.7 $\times$   39.6  &  0.47 $\times$  0.47  &   $-$71  &  0.23  &   867  &   11.0  &   23.6  \\
4 & 16:00:30.5 & $-$53:12:21 &            2.356  &    2.382  &   21.8 $\times$   21.6  &  0.20 $\times$  0.20  &    78  &  0.10  &   121  &    5.2  &   41.7  \\  
5 & 16:01:34.2 & $-$53:11:11 &            2.308  &    4.445  &   33.9 $\times$   26.5  &  0.37 $\times$  0.22  &  $-$103  &  0.14  &   227  &    5.1  &   28.3  \\
6 & 16:00:32.5 & $-$53:12:56 &            2.261  &    2.284  &   21.8 $\times$   21.6  &  0.20 $\times$  0.20  &    34  &  0.10  &   116  &    5.0  &   40.0  \\  
7 & 16:01:07.1 & $-$53:16:28 &            1.950  &    2.798  &   31.0 $\times$   21.6  &  0.31 $\times$  0.20  &    26  &  0.13  &   143  &    4.3  &   25.3  \\
\hline        
 
\end{longtable}
\end{longtab}
}

     {\it Clumpfind}, unlike {\it Gaussclumps}, needs 
     only two parameters (threshold and stepsize) to identify {\it clumps}. 
     The program {\it clfin2d}\footnote{http://www.ifa.hawaii.edu/users/jpw/clumpfind.shtml} 
     is a modification of the original code for three-dimensional
     datacubes. We start the contouring at 3$\sigma_{\rm rms}$ (threshold) with an 
     interval of 2$\sigma_{\rm rms}$ (stepsize) to process each dust emission map. As \cite{pi09} have found, the number of identified sources depends
on the combination of chosen threshold and stepsize. As the stepsize decreases,
the number of clumps increases.

 All the emission is assigned to {\it clumps} above the given threshold.
     It is important to point out that {\it Clumpfind} does not  
     assume Gaussian shape and does not allow overlapping of identified
     sources.
Sources identified by {\it Clumpfind} are listed in Table~\ref{Tclumpfind}. 
     Columns are (1) running number in the order {\it Clumpfind} finds the 
     source, (2)--(3) J2000 position, (4) peak intensity, (5) flux density,
     (6) deconvolved effective radius, (7)--(9) total mass,
     beam-averaged H$_2$ column density, and volume density see Sect.~\ref{estimates}.
     IRDCs G329, G331, G335, G337, G343, and G345 were 
     decomposed into 39, 33, 84, 53, 63, and 80 sources, respectively,
     in total, 352 sources.

\onllongtab{
\begin{longtable}{r c c r r c r r r}
  \caption{Sources found with {\it Clumpfind} and derived physical parameters.}
\label{Tclumpfind}\\
\hline
\hline
\noalign{\smallskip}

$N_{\rm cl}$   &  \multicolumn{2}{c}{Position} &   $I_{\rm peak}$  & $S$ &
$R_{\rm eff}$  & $M$  & $N_{\rm H_2}$ & $n_{\rm H_2}$\\

   & R.A.(J2000)   &  Decl.(J2000)   & (Jy/beam) & (Jy) &
 (pc)  & (\msun) & ($\times 10^{22}~\cmmtw$) & ($\times 10^4~\cmmth$)  \\

\endfirsthead
\caption{continued.}\\
\hline\hline
\noalign{\smallskip}

$N_{\rm cl}$   &  \multicolumn{2}{c}{Position} &   $I_{\rm peak}$  & $S$ &
$R_{\rm eff}$  & $M$  & $N_{\rm H_2}$ & $n_{\rm H_2}$\\

   & R.A.(J2000)   &  Decl.(J2000)   & (Jy/beam) & (Jy) &
 (pc)  & (\msun) & ($\times 10^{22}~\cmmtw$) & ($\times 10^4~\cmmth$)  \\

\hline
\endhead
\hline

\noalign{\smallskip}

\multicolumn{9}{c}{\bf IRDC G329} \\

1 & 16:00:31.2 & $-$53:11:35 &         12.422  &   36.836  &  0.75  &  1884  &   27.2  &    1.6  \\
2 & 16:00:32.5 & $-$53:11:11 &         12.251  &   33.915  &  0.92  &  1734  &   26.9  &    0.8  \\
3 & 16:01:47.7 & $-$53:12:18 &          8.914  &   26.139  &  0.83  &  1336  &   19.5  &    0.8  \\
4 & 16:01:10.3 & $-$53:07:57 &          5.083  &   24.868  &  0.94  &  1271  &   11.1  &    0.5  \\
5 & 16:01:34.2 & $-$53:12:48 &          2.325  &    5.511  &  0.59  &   281  &    5.1  &    0.5  \\
6 & 16:01:55.7 & $-$53:09:03 &          0.704  &    1.454  &  0.42  &    74  &    1.5  &    0.4  \\
7 & 16:01:43.4 & $-$53:05:19 &          0.565  &    0.696  &  0.27  &    35  &    1.2  &    0.6  \\
\hline  

\end{longtable}
}

We do not find a one-to-one correspondence for
all {\it Gaussclumps} and {\it Clumpfind} sources. In each cloud,
{\it Gaussclumps} decomposes emission into more and smaller sources than {\it
  Clumpfind} does, especially around very bright dust peaks.

\subsection{Gas and virial mass, column density, and volume density estimates}
\label{estimates}

    For each identified source and for both methods, we estimated the gas mass
    ($M$), assuming that the submillimeter emission is optically thin, according to
     the expression \citep{hi83}
    \begin{equation}
      M = \frac{S_{\nu}~D^2~R_{\rm gd}}{\kappa_{\nu}~B_{\nu}(T_{\rm d})}\,,
\label{emass}
   \end{equation}
 where $S_{\nu}$ is the observed integrated flux density, $D$  the
    distance, $R_{\rm gd}$ the gas-to-dust mass ratio, $\kappa_{\nu}$ 
     the dust opacity coefficient, and $B_{\nu}$($T_{\rm d}$)  the 
    Planck function at the dust temperature ($T_{\rm d}$).
We assume a gas-to-dust mass ratio of 100 and adopt a 
    $\kappa_{\nu} =$ 1.95~cm$^2$~g$^{-1}$ 
    \citep[interpolated to 870~$\mu$m from Table 1, Col. 9 of][]{os94}, for
    an MRN \citep{mnr}
    graphite-silicate grain mixture with thick ice mantles, 
    at a gas density of 10$^{6}$~\cmmth. 
        In the case of the IRDC G343, whose rms noise is the lowest, and assuming  $T_{\mathrm{d}}$ = 18~K, 
        the detection limit is $\sim$4~\msun~with $S_{\nu}$ = 0.084 Jy (3$\sigma_{\rm rms}$ detection 
    level), at a distance of 2.7~kpc.

 We use $T_{\mathrm{d}}$ = 18~K to compute the masses presented in
 Tables~\ref{Tgaussclumps} and~\ref{Tclumpfind}. This temperature is
 in accordance with previous works toward IRDCs 
\citep[e.g.,][]{pi06a,ra10,mi12}. 
\citet{ra10}
estimated dust temperatures for 190 cores within IRDCs by
 doing graybody fits to their spectral energy distributions (SEDs).
 They obtain median values that range between 23.7$\pm$5.3 and
 40.4$\pm$5.7~K. We are aware that our assumption of low temperature
 is not valid for sources that are associated with masers,
 \hii~regions,~\uchii~regions, and/or IRAS/MSX/24 $\mu$m point sources,
namely regions with signs of active star formation 
 (see Sect.~\ref{laboca:source-class},) whose temperature should be
 higher.  On the other hand, sources not associated with any of those
 star formation signposts are expected to have a lower temperature.  
We note that by
 decreasing the temperature from 18~K to 12~K, the mass would
 almost double, while with an increase from 18~K to 30~K,
 the mass would decrease by 50\%.

      Using the N$_2$H$^+$~(3--2) FWHM line widths ($\Delta{\rm V}$), we 
estimate the virial masses ($M_{\rm vir}$)
      for 11 positions. The virial mass of a clump with a constant density
      distribution  is expressed by
  $M_{\rm vir}~\simeq~210 \times \left(\frac{R}{\rm pc}\right)~\left(\frac{\Delta{\rm V}}{\rm km~s^{-1}}\right)^2$ \citep{mac88},
 where  $R$ is the clump radius. We assumed $R$ as equal to the effective radius
($R_{\rm eff} = \sqrt{A/\pi}$, with $A$ the area of the source). 
If, on the other hand, the density structure varies
as $\rho \propto 1/r^2$, the vivial mass is 
$M_{\rm vir}~\simeq~126 \times \left(\frac{R}{\rm pc}\right)~\left(\frac{\Delta{\rm V}}{\rm km~s^{-1}}\right)^2$. 
The results are listed in Table~\ref{Tn2hp:laboca}.
The virial parameter  \citep{ber92} defined as 
$\alpha_{\rm vir} \equiv M_{\rm vir}/M_{\rm tot}$
has a mean and standard deviation of 1.6 and 0.94, respectively, for a constant
density distribution, while the values are 0.95 and 0.56 for 
$\rho \propto 1/r^2$.

The beam-averaged column density, $N_{\rm H_2}$, is computed using the
    expression in
\begin{equation}
      N_{\rm H_2} = \frac{I_{\nu}^{\rm peak}~R_{\rm gd}}{\kappa_{\nu}~B_{\nu}(T_{\rm d})~\Omega~\mu_{\rm H_2}~m_{\rm H}}\,,
\label{enh2}
\end{equation}
     where $I_{\nu}^{\rm peak}$  is the peak intensity,
$\mu_{\rm H_2}$   the molecular weight per hydrogen molecule, and
$\Omega$   the beam solid angle.
 We adopt $\mu_{\rm H_2} =$ 2.8 \citep{ka08} and the definition of 
   $\Omega = (\pi \theta_{\rm HPBW}^2)/(4 \rm{ln2}) $ with 
      $\theta_{\rm HPBW}$ as the half-power beam width. 
   In the case of the IRDC G343, these observations are sensitive to
   column densities as low as $N_{\rm H_2} = 1.8~\times 10^{21}$~\cmmtw~with 
   $I_{\nu}^{\rm peak}$~=~0.084~Jy~beam$^{-1}$ (3$\sigma_{\rm rms}$ detection level)
   and $T_{\mathrm{d}}$ = 18 K.

 Volume densities are computed, assuming spherical configuration for the 
   identified sources, as 
\begin{equation}
n_{H_2} = \frac{M_g}{\frac{4}{3}\pi R_{\rm eff}^3\mu_{H_2} m_H}\,,
\end{equation}
   where $R_{\rm eff} = \sqrt{A/\pi}$ is the effective radius and
   $A$ is the area of the source.

We see that {\it Gaussclumps} tends to decompose the emission into  
   smaller sources than {\it Clumpfind}. The mean, minimum, and maximum 
$R_{\rm eff}$  for the {\it Gaussclumps} method, are 0.20, 0.09, and 
0.74~pc, respectively, taking  a total of 
   510 sources into account. The {\it Clumpfind} mean, minimum, and maximum 
$R_{\rm eff}$ are 0.40, 0.16, and 0.99~pc, respectively, for  
   352 sources. Given these sizes, the sources we discuss in this
   contribution are considered as clumps.

As for the clump masses, we get a mean, minimum, and maximum mass of 111,
 6, and  2692~\msun, respectively, with {\it Gaussclumps},  while
   for {\it Clumpfind} the values are 141, 7, and 4254~\msun. 
   The total mass of the decomposed clumps, taking into account all  
   clouds, is 56\,587~\msun~for 
   {\it Gaussclumps} and 49\,722~\msun~for {\it Clumpfind}.

\subsection{Star formation signposts and clustering}\label{laboca:source-class}

\begin{figure*}
   \centering
\includegraphics[width=7.0cm]{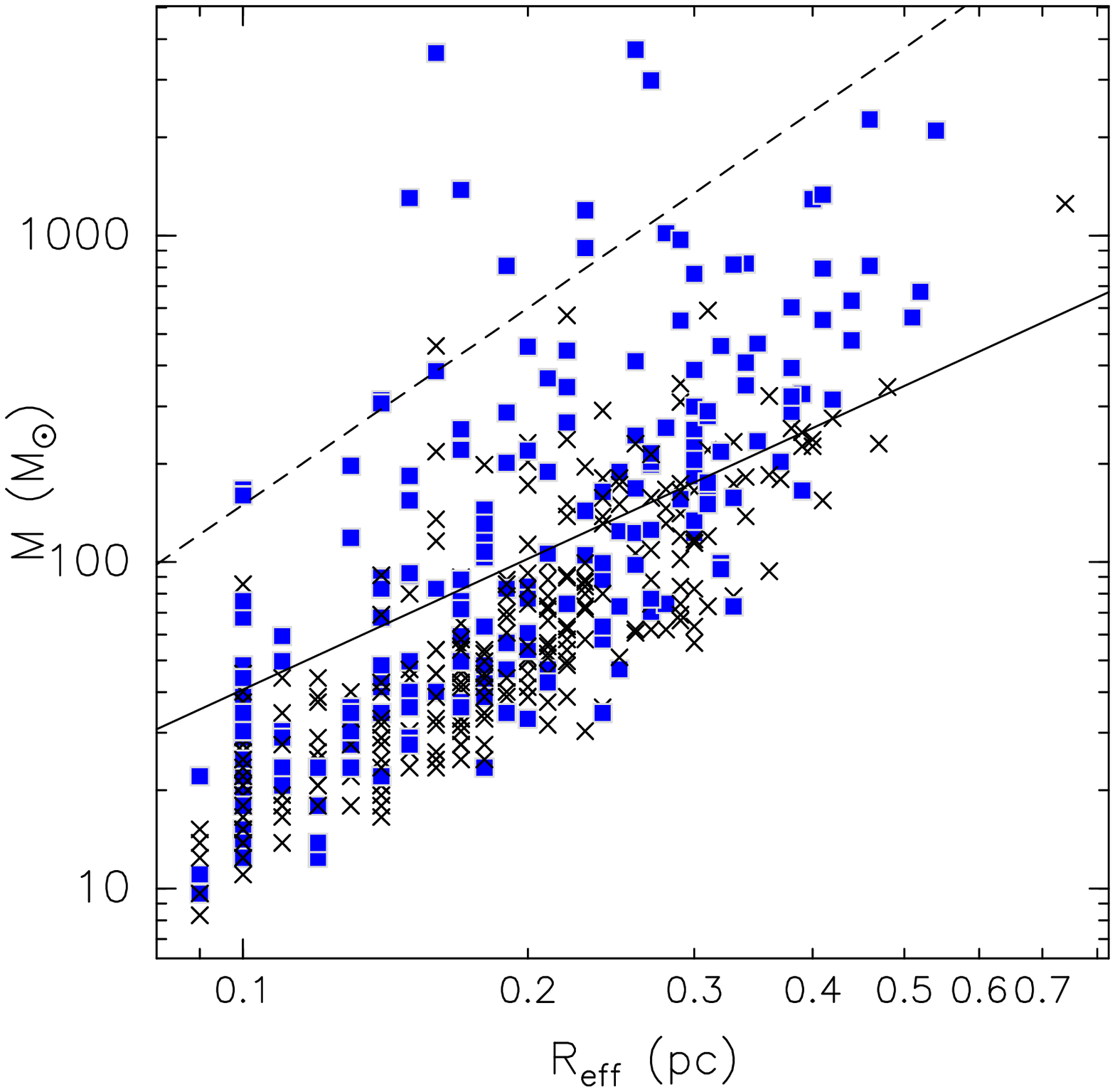}\hspace{0.2cm}
\includegraphics[width=7.0cm]{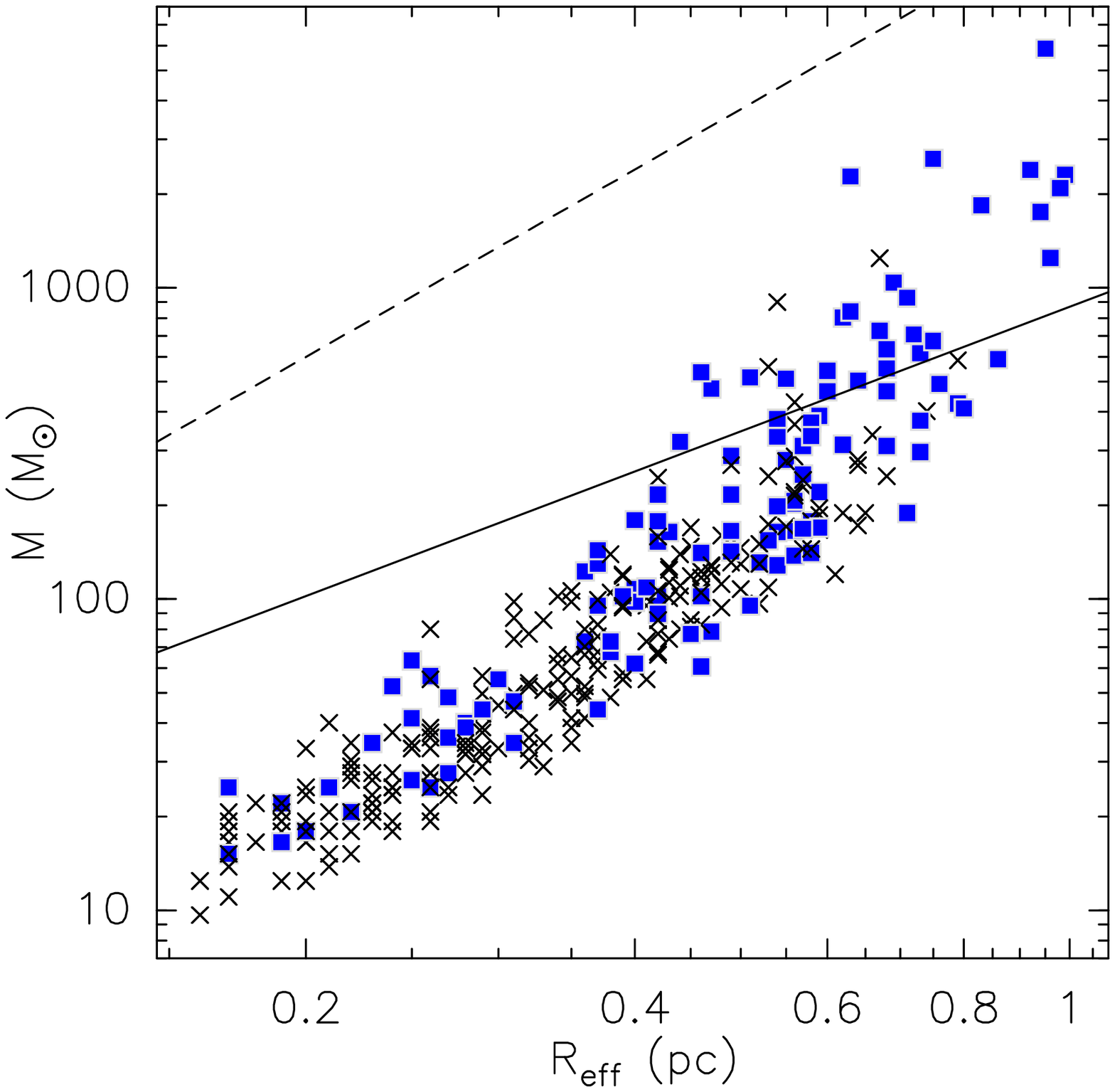}
\caption{
  Mass-size plots for clumps extracted with {\it Gaussclumps} 
(left panel) and {\it Clumpfind} (right panel) for all clumps in our
  IRDCs. The solid lines represent the empirical lower limit
  threshold for high-mass star formation 
$M \propto R_{\rm    eff}^{1.33}$ found by \citet{ka10}, while the dashed lines
  represent the theoretical minimum threshold of 1~g~\cmmtw~ proposed
  by \citet{kru08}.  Blue filled squares represent clumps associated
  with any of the following signposts of star formation:
  IRAS/MSX/24-$\mu$m point sources, masers, \hii~regions, and/or
  \uchii~regions. Crosses represent clumps with no reported  signpost
  of star formation. 
}
\label{Fmass-size}
\end{figure*}

   We cross-identified the {\it Gaussclumps} and {\it Clumpfind} sources with signposts of star
   formation, such as IRAS/MSX point sources, masers (H$_2$O, CH$_3$OH, OH),
   green extended objects \citep[EGOs, i.e., shocked regions;][]{cy08},
   \hii~regions, and~\uchii~regions using the Set of Identifications, Measurements, and
   Bibliography for Astronomical Data (SIMBAD4, release 1.181) as of 
   July 2011. Point sources at 24 $\mu$m were identified by visual 
   inspection of {\it Spitzer}/MIPSGAL maps.
   Only signposts that are in the 
   {\it FWHM} ellipse of {\it Gaussclumps} sources or within the limits of
{\it Clumpfind} sources are considered and marked 
   accordingly in Fig.~\ref{Fmass-size}. 
Statistics of these
   identifications are shown in Table~\ref{Tstats}.

   The clustering per IRDC that was  measured 
   with the mean clump density parameter is listed in Table~\ref{Tstats}. The mean clump density ranges from 
   0.11--0.26 arcmin$^{-2}$ for {\it Gaussclumps} and 0.09--0.18 for {\it Clumpfind}.

\begin{table*}
\begin{minipage}{\textwidth}
  \caption{Statistics of cross-identifications of {\it Gaussclumps}/{\it Clumpfind} sources with star formation signposts and clustering information.}
\label{Tstats}
\centering
\renewcommand{\footnoterule}{}
    \begin{tabular}{l c c c c c c}
\hline
\hline
\noalign{\smallskip}

SF signpost  & \multicolumn{6}{c}{IRDC} \\
            & G329  & G331  & G335  &  G337  & G343 & G345 \\
\hline

IRAS point source (\%) & 5/10 & 2/6 & 7/8 & 2/4 & 1/3  &  4/8  \\
MSX point source(\%)   & 3/8  & -/-   & 1/1 & 2/2 & -/-    &  2/1 \\
24 $\mu$m-bright(\%)   & 40/28  & 46/46  & 37/36  & 28/36 & 34/27 &  46/26 \\
\hii/\uchii (\%)       & -/-   & 5/6 & -/-   & 2/2 & -/-    &  2/3  \\
Masers     (\%)        & 13/10  & 2/3 & 1/1 & -/-   & -/-    &  2/1 \\
EGO         (\%)       & 8/10   & -/-   & 2/2 & 2/2 & 7/6  &  2/3 \\

\hline
\hline
\noalign{\smallskip}
Clustering        &     &    &      &     &      &      \\
\noalign{\smallskip}
\hline
\noalign{\smallskip}

Number of clumps  &   75/39   & 41/33  &  123/84  & 65/53  &  83/63  &  123/80   \\
Area (arcmin$^2$) &  342   & 358   &   475   &  449   &  561    &  640    \\
Mean clump density (arcmin$^{-2}$)&  0.22/0.11  & 0.11/0.09  & 0.26/0.18  &  0.14/0.12   &  0.15/0.14   &   0.20/0.13   \\

\hline  
\end{tabular} 
\tablefoot{SIMBAD4 was used for performing cross-identifications. EGO stands for
extended green object at 4.5 $\mu$m \citep[see][]{cy08}.}
\end{minipage}
\end{table*}

\subsection{Comments on individual IRDCs (based on {\it Gaussclumps})}\label{comments}

{\bf IRDC G329}. The IRAS sources 15566-5304, 15579-5303, 15573-5307, and 15574-5306 are 
associated with  clumps in the field of G329. The EGOs identified in this 
region were cataloged as ``possible'' massive young stellar object (MYSO) 
outflow
candidates by \citet{cy08}. 
Masers (OH and CH$_3$OH) toward a dozen of clumps were
reported by \citet{ca95}.

{\bf IRDC G331}. The source IRAS~16070$-$5107 is associated with a  clump in the
field of G331. One clump is associated with the radio continuum source 
G331.4+00.0, which
was observed in an all-sky survey of \hii~regions at 4.85 GHz \citep{ku97}.
 We retrieved C$^{18}$O (1--0) data from the Three-mm Ultimate Mopra Milky way 
Survey 
(ThrUMMS\footnote{http://www.astro.ufl.edu/~peterb/research/thrumms/}) 
and found that the emission in the lower left-hand corner in G331  
(see Fig.~\ref{Firdcslaboca}), where the radio source is located, has
a different systemic velocity ($\sim -82$~\kms).

{\bf IRDC G335}. Together with the IRDC G343, G335 has the highest number of clumps. Eight
IRAS sources are associated with clumps in G335. We find two EGOs in this
region: the first one, G335.43$-$0.24, was cataloged as a ``possible'' MYSO 
candidate and the second one, G335.06$-$0.43, as a ``likely'' MYSO outflow candidate
\citep{cy08}. The bubble S42 was identified by visual inspection of 
{\it Spitzer}/GLIMPSE images in \citet{ch06} and was cataloged as a
``broken or incomplete'' ring. The emission at 870 $\mu$m
is tracing the eastern side of the dusty shell
(see the {\it bottom} panel of Fig.~\ref{Firdcslaboca}).

{\bf IRDC G337}. One clump is associated with the source IRAS~16340-4732 .
The EGO G337.16$-$0.39  is found in this field and was cataloged a ``likely'' 
MYSO outflow candidate \citep{cy08}.
The Galactic radio source GRS~337.10$-$00.20, located northwest
of the central filament, is probably not part of the IRDC complex
because of different velocity components found for several line tracers.
Previous studies of 337.10$-$00.20 show two velocity components in the 
H90$\alpha$ profile, at $-73$ \kms~and at $-59$ \kms~\citep{sa97}, one velocity
component at $-73$ \kms~in the H190$\alpha$ profile \citep{wi70}, and four
velocity components in the C{\scriptsize I} profile, at 108, $-75$, $-36$, and 
$-22$ \kms~\citep{hu99}. We confirm that this part of the field is at 
a different systemic velocity ($\sim -72$~\kms) by retrieving N$_2$H$^+$ (1--0)
data from the Millimeter Astronomy Legacy Team 90~GHz (MALT90) survey \cite[][]{fo11}.

{\bf IRDC G343}. The source IRAS~16575-4252 is associated with clumps in the field of G343.
Three ``possible'' MYSO outflow candidates  and one ``likely'' outflow candidate 
from \citet{cy08} are 
associated with clumps in this region. No maser observations are reported in the
literature.

{\bf IRDC G345}. The three IRAS sources 17010-4124, 17014-4129, and 17018-4127 are associated 
with clumps in this field. Masers (OH, H$_2$O, and CH$_3$OH) were
reported by \citet{ca95} toward a couple of clumps.
The EGOs  G345.00-0.22a, G345.00-0.22b, and G345.13-0.17 were
cataloged by \citet{cy08} as ``possible'' MYSO outflow candidates.
\cite{fi03} have found a distance of 2.9 kpc to one of the clumps associated
with an IRAS source,
which is actually a UC \hii~\citep{ga06}. \citet{ga07} carried out
observations with the Swedish-ESO Submillimeter Telescope (SEST) at 1.2 mm 
toward the southern part of the filament that connects to the 
UC~\hii~G345.001-0.22. 
The morphology at 1.2~mm is similar to that we see 
in the {\it bottom} panel in Fig.~\ref{Firdcslaboca2}. 
These authors used a kinematic distance of 2.7~kpc. We use a distance of 
2.8~kpc based on our 
N$_2$H$^+$ (3--2) line observations.

\subsection{Mass-size relation for HMSF}\label{laboca:mass-size}

   Figure~\ref{Fmass-size} shows the mass-size relationship for
   clumps extracted with {\it Gaussclumps} (left panel) and {\it
     Clumpfind} (right panel). We plot two HMSF lower thresholds: the
   one proposed by \citeauthor*{kru08} (2008; hereafter \citetalias{kru08}), 
   who found a limit
   in $N_{\rm H_2}$ = $2.13 \times 10^{23}$ \cmmtw~(or 1 g~\cmmtw) to
   avoid fragmentation and to allow high-mass stars to form, and
   the one discussed on empirical basis by \citeauthor*{ka10} (2010; 
   hereafter \citetalias{ka10}).
    For the sake of comparison with those criteria, clump
   masses were computed againby decreasing the dust opacity values given
   in \citet{os94} by a factor of 1.5, as done in KP10. Additionally, we
   make a reduction of $\ln(2) \approx 0.69$ in the total mass to
   account for the mass contained in the half peak column density
   contour.

In Fig.~\ref{Fmass-size}, 
the percentage of clumps (found with {\it Gaussclumps}) that lie 
        above the \citetalias{ka10} relation with and without association to star formation
        signpost is 19\% and 9\%, respectively, while 3\% and $<$1\%
        of clumps (with and without association to star formation
        signpost) satisfy the much more stringent threshold of 
        \citetalias{kru08}.
        The percentages of clumps (identified with {\it Clumpfind})
        above the  \citetalias{ka10} relation are 8\% and 1\% (with and 
        without association to star formation
        signpost). All  {\it Clumpfind} sources lie below the threshold of 
        \citetalias{kru08}.

\subsection{Mass distribution of clumps in IRDCs}

\begin{figure*}
   \centering
   \includegraphics[width=6cm]{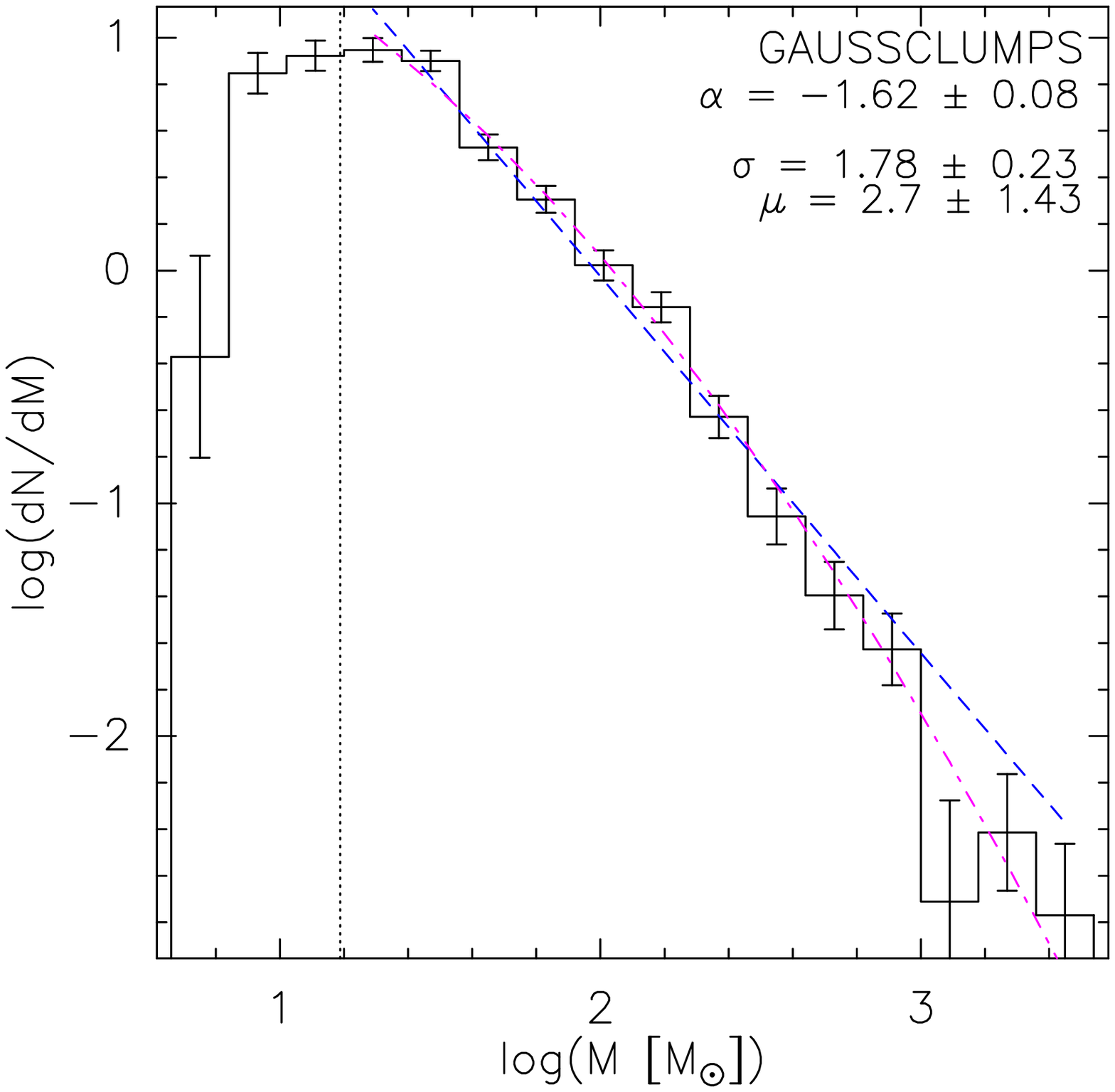}\hspace{0.1cm}
   \includegraphics[width=6cm]{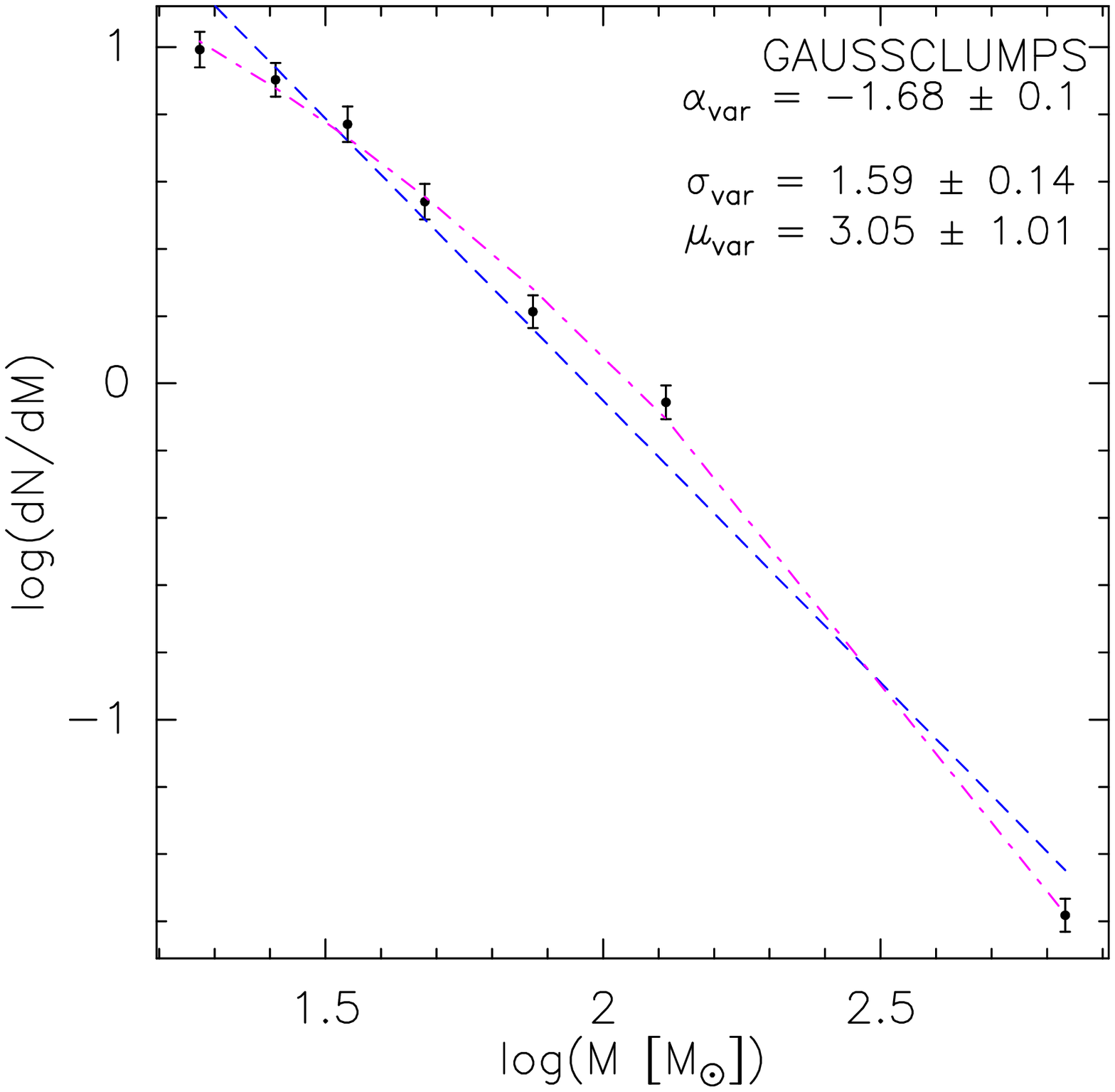}\\ \vspace{0.1cm}
   \includegraphics[width=6cm]{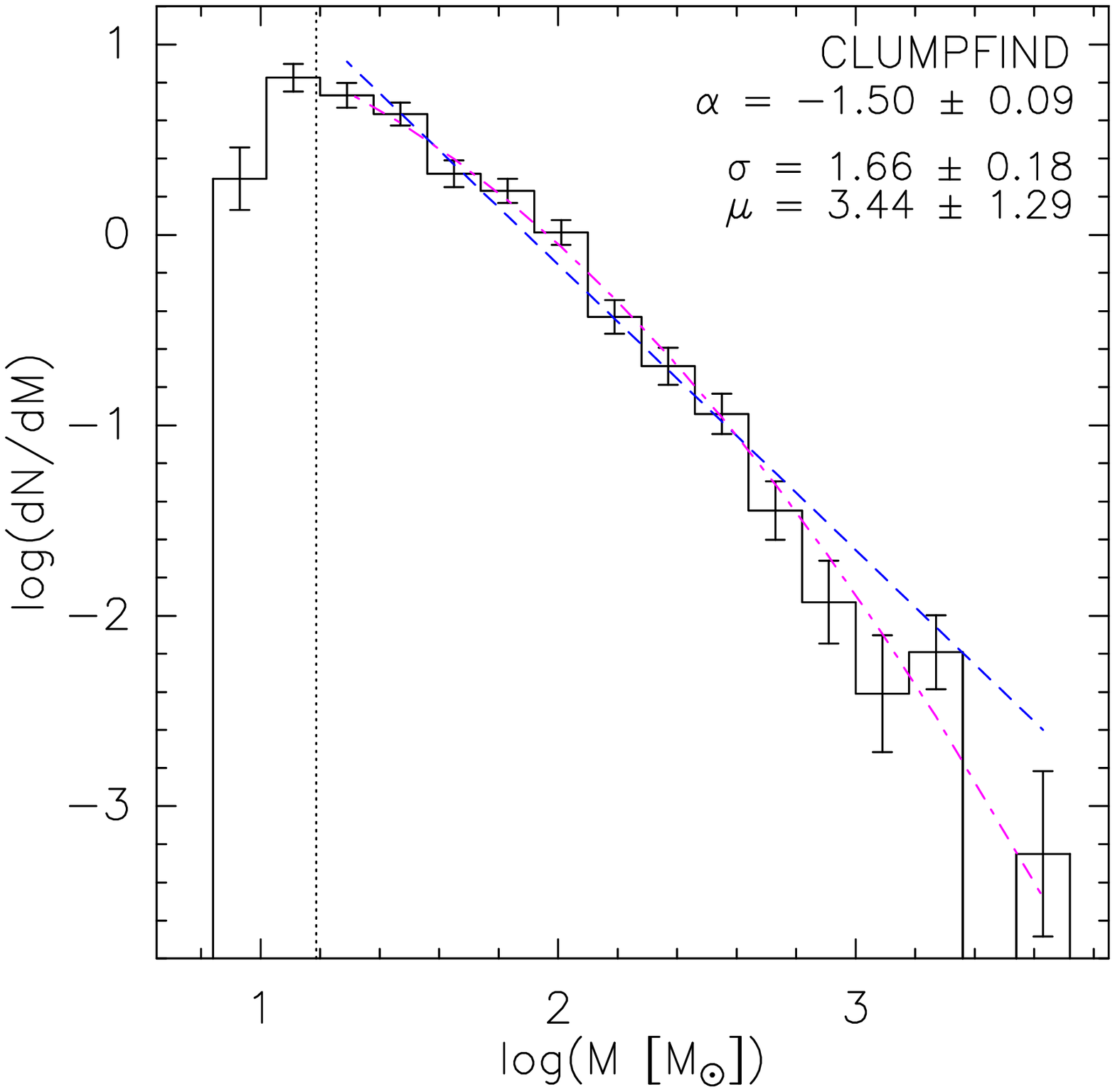}\hspace{0.1cm}
   \includegraphics[width=6cm]{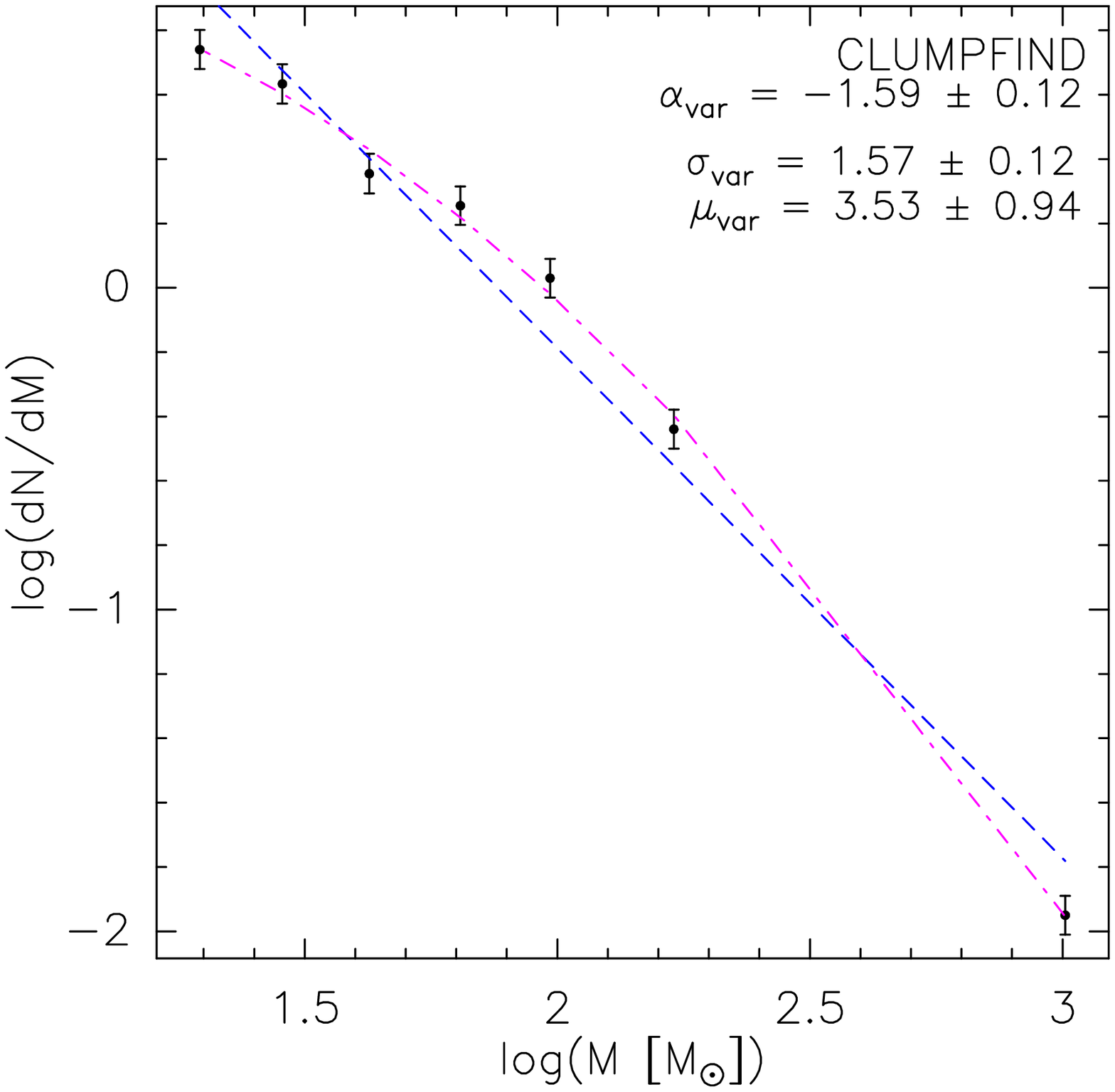}
\caption{ Differential mass functions, $dN/dM \propto M^{\alpha}$, for
  clumps from {\it Gaussclumps} ({\it upper} panels) and clumps from
  {\it Clumpfind} ({\it lower} panels). The dashed blue lines
  represent fits to single power laws 
and the dot-dashed pink lines represent the fits
  to log-normal distributions.
The
 parameters
$\alpha$, $\sigma$, and $\mu$ are given in each panel.
The vertical dotted lines
  indicate the 6$\sigma_{\rm rms}$ (15~\msun) mass given by the
  noisiest map.}
\label{Falldndm-gcl-clf}
\end{figure*}

In Fig.~\ref{Falldndm-gcl-clf}, we present the differential mass
functions for the 510 clumps extracted with {\it Gaussclumps} (upper
panels) and the 352 sources from {\it Clumpfind} (lower panels).  We
plot the mass distribution using two approaches: in the first (left
panels), the bin size was uniform with uncertainty given by a Poisson
distribution.  In the second (right panels), we followed the technique
by \citet{ma05} in which the bin size is variable so that the number
of clumps per bin is approximately constant in order to minimize the
binning biases. The uncertainty is derived from a binomial
distribution according to \citet[][]{ma05}.

   In all cases we fit the differential mass distribution with a
   single power-law function
\begin{equation}
\frac{dN}{dM} \propto M^{\alpha}\,,\label{Esinglepl}
\end{equation}
  where $dN$ is the number of objects in $dM$, $dM$  the mass bin,
  and $\alpha$ the power-law index.  
 In addition, we fit a lognormal function:
\begin{equation}
\frac{dN}{dM} \propto \frac{1}{M\sigma}\exp\left[-\frac{(\ln M -
    \mu)^2}{2\sigma^2}\right]\,,\label{Elognor}
\end{equation} 
where $\sigma$ is the dispersion and $\mu$ is related to the peak mass
($M_{\rm peak}$ = $e^{\mu-\sigma^2}$).

In Fig.~\ref{Falldndm-gcl-clf}, we present
the least-squares fits to single power laws, $dN/dM \propto M^{\alpha}$, 
with $\alpha = -1.62\pm0.08$ and 
$\alpha_{\rm var} = -1.68\pm0.10$ for {\it Gaussclumps} ({\it upper} panels) 
with uniform and variable bin size, respectively, and to slopes 
$\alpha = -1.50\pm0.09$ and $\alpha_{\rm var} = -1.59\pm0.12$ for 
{\it  Clumpfind} ({\it lower} panels).  
In Fig.~\ref{Falldndm-gcl-clf}, we also present
fits to log-normal distributions, with 
$M_{\rm peak} = 0.61\pm0.92$ \msun, $\sigma = 1.78\pm0.23$ and 
$M_{\rm peak,var} = 1.70\pm1.76$ \msun, $\sigma = 1.78\pm0.23$ for 
{\it Gaussclumps} and $M_{\rm peak} = 1.98\pm2.64$ \msun, 
$\sigma = 1.66\pm0.18$ and $M_{\rm  peak,var} = 2.91\pm2.79$ \msun, 
$\sigma = 1.57\pm0.12$ for {\it  Clumpfind}.  The vertical dotted line 
indicates the 6$\sigma_{\rm  rms}$ (15~\msun) mass given by the noisiest map.  
As for the variable bin size, masses higher than 15~\msun~were plotted and used
in the fit. Since the fits to log-normal functions result in peak
masses, $M_{\rm peak}$, below our 6$\sigma_{\rm rms}$ threshold of
15~\msun, in what follows, we focus on the index obtained with
the single power-law fits.

     We find good agreement, within the uncertainties, between the mass 
     distribution obtained with either {\it Gaussclumps} or {\it Clumpfind}. 
     Moreover, these indices were similar when using a variable 
     or uniform bin size.  The single power-law index of our IRDC mass 
     distribution has a mean and standard deviation of 
     $\alpha =-1.60$ and 0.06, respectively.

To check the effects in the 
mass distribution index due to different
assumptions in, say, the source decomposition, estimation of masses,
 contribution of extended emission, and  temperatures, we 
tested several scenarios with {\it Clumpfind}, using a 3$\sigma$ threshold and
a 2$\sigma$ stepsize (see Appendix~\ref{appendix1} for more details).
The obtained indices have a mean and standard deviation of
 $\alpha = -1.68$ and 0.15, respectively.


\section{Discussion}\label{laboca:dis}

\subsection{Criteria for high-mass star formation}

        Less than 19\% of the clumps found with {\it Gaussclumps} 
        lie above the \citetalias{kru08} and \citetalias{ka10} thresholds. 
        The majority of them have the mass 
        needed to form high-mass stars, or they have already formed them. 
        Clumps with no association to a star formation signpost are very 
        interesting candidates for sources 
        in the earliest phases of high-mass stellar cluster evolution. 
        All clumps found with {\it Clumpfind} are located 
        below the
        \citetalias{kru08} threshold. However, some of these clumps, as well
        as some found with {\it Gaussclumps} may contain
        higher surface density structures that are diluted within the beam.
        This can also explain the difference in percentage of clumps
        above both thresholds using different decomposition methods 
        (see Fig.~\ref{Fmass-size}). As we found in 
        Sect.~\ref{laboca:decomposition}, {\it Gaussclumps}
        tends to find more and smaller sources than does
        {\it Clumpfind}.

        On the other hand, the eleven clumps observed in N$_2$H$^+$ (3--2)
        are all above the threshold
        discussed by \citetalias{ka10}, while five of them lie above the 
        \citetalias{kru08} one. 
        Additionally, we can consider these clumps as dominated by gravity
        and either on the verge of collapse or already collapsing sources,
        according to their virial parameters.

\subsection{Mass spectra}

 Making a comparison of the mass distribution indices is a hard task due to the
many assumptions authors made when,  extracting sources and estimating 
masses, etc. Several authors use extinction maps \citep[][]{si06,ma09,pe10b,ra09} 
while others make use of (sub)millimeter continuum maps
\citep[][]{ra06,mi12}.
In an attempt to compare our results with those of other authors, we
 fitted a single power law to our mass distribution using as many of their 
assumptions as possible: e.g., temperature, dust opacity, and 
threshold and stepsize (in the case of {\it Clumpfind}).

\begin{figure*}
\centering
\includegraphics[scale=0.45]{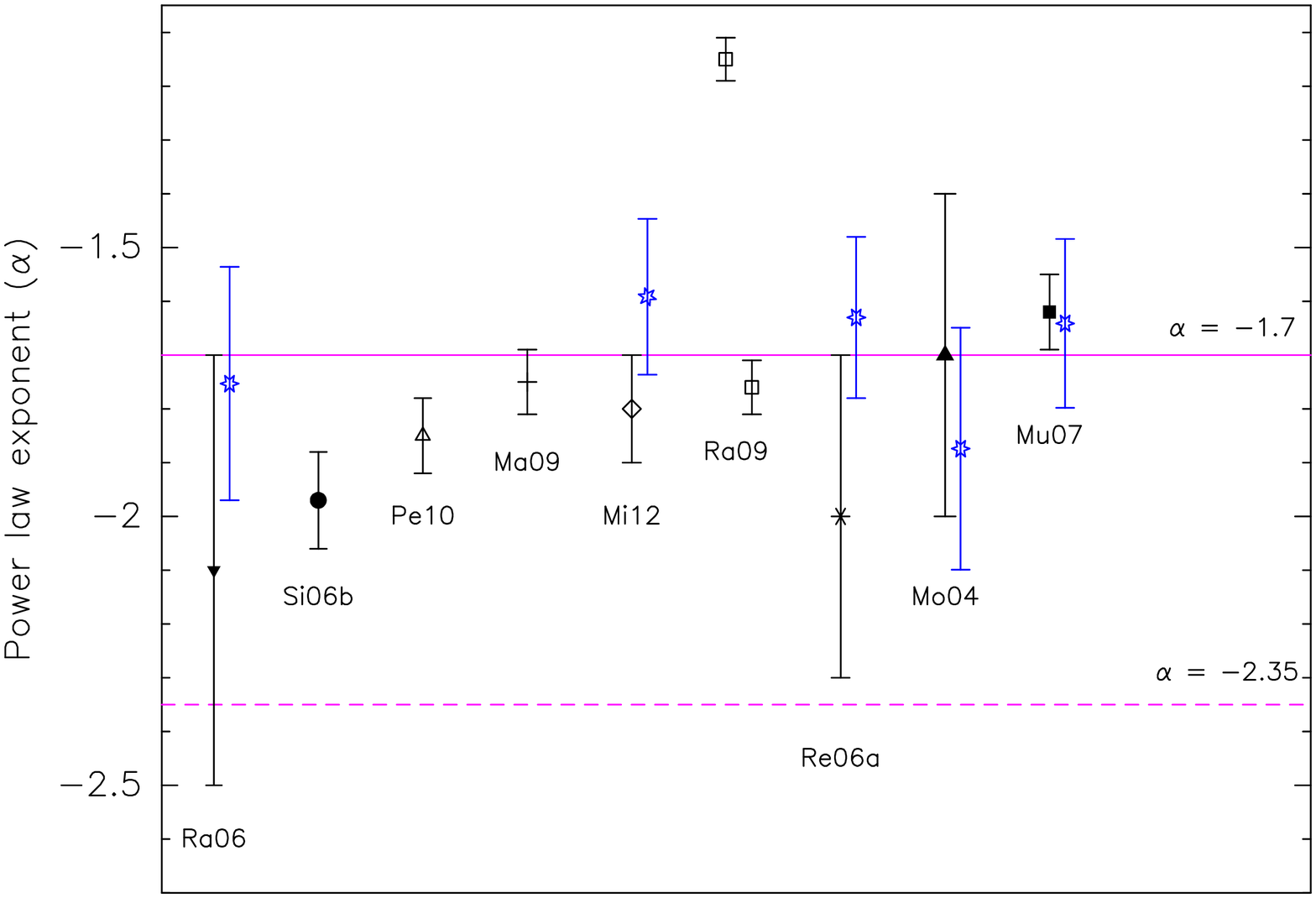}
\caption{Comparison of $\alpha$ values from the literature and our estimates.
All symbols but six-pointed stars represent values obtained
by other authors. Six-pointed star symbols represent our estimations taking 
different assumptions into account. Vertical lines represent the 1$\sigma$
uncertainties. Horizontal lines show
the index of the CO clump 
mass distribution (solid line) and the index 
of the IMF (broken line). References:  Ra06 \citep{ra06}, Si06b \citep{si06b}, 
Pe10 \citep{pe10b}, Ma09 \citep{ma09}, Mi12 \citep{mi12}, Ra09 \citep{ra09}, 
Re06a \citep{re06}, Mo04 \citep{mo04}, Mu07 \citep{mu07}.}
\label{Findices}
\end{figure*}

 Figure~\ref{Findices} present the indices of the cloud/clump/core mass
distribution for IRDCs and other high-mass star-forming regions from
the literature.
 We  plot  those indices and compare them with our estimated values.

  \citet{ma09}, \citet{ra09}, and~\citet{pe10b} 
have studied the mass distribution of 
  IRDCs obtained from extinction maps.   
In particular, \citet{ra09} computed the mass distribution of cores
using {\it Clumpfind} in 11 IRDCs and fitted broken power laws of
$\alpha_{\rm low} = -0.52 \pm 0.04$ and $\alpha_{\rm high} = -1.76 \pm 0.05$  
  for masses lower than 40~\msun~and for masses greater
than 40~\msun, respectively. Using {\it Gaussclumps}, the slope
becomes shallower: $\alpha_{\rm high} = -1.15\pm 0.04$ and 
$\alpha_{\rm  low} = -0.64\pm 0.07$  for 
the same break-point mass.  We plot in Fig.~\ref{Findices} $\alpha_{\rm high}$
for both {\it Gaussclumps} and {\it Clumpfind} since they are not
consistent with each other within the uncertainties. 
\citet{pe10b} find that a
lognormal distribution better fitted the mass distribution of
``fragments''
with masses higher than 10~$M_\odot$. They also obtained
the index for clouds, $\alpha = -1.85 \pm 0.07$. 
\citet{ma09} found a similar $\alpha = -1.75 \pm 0.06$ for 
$M_{\rm clouds} > 1.7 \times 10^3~\msun$. 
     
     Using $^{13}$CO observations, \citet{si06b} find a mass
     distribution of $\alpha = -1.97 \pm 0.09$ for IRDCs.  This slope
     results from a fit to the high-mass end of the molecular cloud
     distribution (for $M_{\rm clouds} > 10^{3.5}~\msun$). 

 We computed our $\alpha$ value again to compare it with      
the $\alpha = -2.1
     \pm 0.4$ estimated by \citet{ra06} for cores within IRDCs.
      They extracted core properties from the 1.2 mm dust continuum
     emission.  We fitted the mass distribution including clumps
     with no signposts of star formation
     (extracted with {\it Gaussclumps}) and used a temperature of 15~K.
      We obtained      $\alpha = -1.75 \pm 0.21$. 
As we can see in Fig.~\ref{Findices}, their result is consistent with our 
estimate, but it is also consistent, 
within their uncertainties,
     with the stellar IMF ($\alpha = -2.35$). 
 
     \citet{mi12} mapped four IRDCs with the LABOCA instrument
     and obtained $\alpha = -1.8 \pm 0.1$ for clump masses above
     1500~\msun.  We ran {\it Clumpfind} using a 3$\sigma$
threshold and a 3$\sigma$ stepsize and used three temperatures: 15~K for the
clumps with no signpost of star formation, 30~K for \hii~regions and 
IRAS sources, and 20~K for the rest of the clumps. 
The bin size was $ \Delta \log (M/\msun) \approx 0.44$. With these
assumptions, we obtained      
$\alpha = -1.59 \pm 0.14$.

   \citet{mo04} studied the giant molecular cloud RCW~106 at 1.2 mm
   and found indices of $\alpha = -1.5 \pm 0.3$ (using {\it
     Gaussclumps}) and $\alpha = -1.7 \pm 0.3$ (with {\it Clumpfind})
   after decomposing the dust emission into clumps. 
    Since their values using two algorithms are
   consistent with each other, in Fig.~\ref{Findices}, we  plot the
 value for {\it Clumpfind}.
   In this case, our estimation of $\alpha = -1.87 \pm 0.22$ was obtained by
   assuming two different temperatures: 20~K for the clumps with no
   signposts of star formation and 40~K for the rest.

   The mass distribution of the HMSFR NGC~6334 has been found to be
   similar to the CO mass distribution with $\alpha = -1.62\pm 0.07$,
   from observations at 1.2 mm \citep{mu07}. We ran {\it Clumpfind}
   using a threshold of 3$\sigma$, a stepsize of 2$\sigma$, and a single temperature of 17~K.
   By making those assumptions, we obtain $\alpha = -1.64\pm 0.16$.

\citet{re05,re06}
   decomposed emission of (sub)mm maps (at 450 $\mu$m and 850 $\mu$m)
   of the HMSFRs NGC 7538 and M17 with {\it Clumpfind}.  Broken power
   laws were fitted to their differential mass functions at 
   870 $\mu$m with indices
   at the high-mass end of $\alpha_{\rm high} = -2.0\pm 0.3$ for
   NGC~7538 and $\alpha_{\rm high} = -1.5\pm 0.1$ for M17.
    In Fig.~\ref{Findices}, we plot the value for NGC~7538.
    Our
    computation of the index assumes a 3$\sigma$ threshold and a 2$\sigma$
    stepsize for {\it Clumpfind},
    a temperature of 35~K, an opacity spectral index of $\beta=1.5$, and 
    $\kappa_\nu =$~0.87~cm$^2$~gr$^{-1}$. We then obtain
    $\alpha = -1.63\pm 0.15$.
   It is important to point out that 
   \citet{re06b} carried out a study in
   which they found that the high-mass end of seven mass
   distributions, including those of NGC~7538 and M17, is 
   $\alpha_{\rm high} = -2.4 \pm 0.1$, resembling that of the stellar IMF.

    In addition, we used the flux densities from the ATLASGAL
     compact source catalog of \cite{co13} for five of our regions in order
   to fit the clump mass distribution. Assuming a single temperature of
   18~K, we found $\alpha = -1.70\pm 0.30$ (for clumps with 
$M_{\rm clumps}>$ 100~\msun), 
which is consistent with the value
   obtained by us.

   As we can see from this revision and from Fig.~\ref{Findices}, 
   our derived $\alpha$'s, as well as most of those obtained by other authors, 
   are consistent with each other within the uncertainties and with the
   CO clump mass distribution.

\section{Summary}\label{laboca:sum}

         We mapped the 870~$\mu$m dust continuum emission of six IRDCs 
         with the LABOCA   instrument  and carried out 
         molecular line observations of the  N$_2$H$^+$ (3--2) line with the
         APEX2A  receiver both with the  APEX telescope.
Our main results can be summarized as follows:
   \begin{itemize}
     \item We  estimated (``near'') kinematic distances of 2.7--3.2~kpc
 using the N$_2$H$^+$ (3--2) line.

 \item We  obtained virial masses for 11 clumps. Their virial
   parameters indicate that
       these clumps are dominated by gravity, either on the verge of
       collapse or already collapsing.

      \item    Each IRDC was decomposed into clumps by using two automated
         algorithms, namely {\it Gaussclumps} and {\it Clumpfind}.
                  The mean $R_{\rm eff}$ is 0.20 pc 
   for the {\it Gaussclumps} method, taking a total of 
   510 sources in account, while the {\it Clumpfind} mean 
$R_{\rm eff}$ is 0.40 pc for  
   352 sources. The clump masses have been found in the range of
6 to 2692~\msun~for {\it Gaussclumps} and 7 to 4254~\msun~for {\it Clumpfind}.

          \item The percentage of clumps that lie 
        above the HMSF threshold discussed by \citet{ka10} with and without 
        association to star formation
        signpost is 19\% and 9\%, respectively, while  3\% and $<$1\%
        of clumps (with and without association to star formation
        signpost) satisfy the much more stringent threshold of \citet{kru08}.
        The percentages of clumps (identified with {\it Clumpfind})
        above the  \citetalias{ka10} relation are 8\% and 1\% (with and 
        without association to star formation
        signpost). All  {\it Clumpfind} sources lie below the threshold of 
        \citetalias{kru08}.

        \item Using two methods of binning, the mass distribution of the 
          decomposed emission into clumps has been fitted with
        a power law whose index is $\alpha =-1.60 \pm 0.06$. This index is
        consistent with the CO clump mass distribution and other high-mass
        star-forming regions.

\end{itemize}

\begin{acknowledgements}
L.~G. acknowledges support for this research from the International Max Planck 
Research School (IMPRS) for Astronomy and
Astrophysics at the Universities of Bonn and Cologne, CONICYT (Chile) 
through project BASAL PFB-06, and CSIRO Astronomy and Space Science. L.~G. would like to thank A. Belloche for his 
help with the use of  {\it Gaussclumps}.  We thank the
 referee for providing helpful comments and suggestions.
This work was partially
funded by the ERC Advanced Investigator Grant GLOSTAR (247078). 
This work is based in part on 
observations made with the {\it Spitzer} Space Telescope, which is operated by 
the Jet Propulsion Laboratory, California Institute of Technology, under a 
contract with NASA. This research made use of SIMBAD database,
operated at the CDS, Strasbourg, France.
\end{acknowledgements}

\bibliographystyle{aa}

\bibliography{labocab}

\Online

\begin{appendix} 
\section{Tests using {\it Clumpfind}}\label{appendix1}

 \citet[][]{pi09} point out several weaknesses in the
{\it Clumpfind} algorithm for estimating the mass distribution
index when choosing the threshold and stepsize parameters.
To check potential assumptions that can affect the estimation of
the clump mass distribution index, we run
{\it Clumpfind} for four different thresholds (2, 3, 4, and 7$\sigma$) and 
nine stepsizes (from 2--10$\sigma$ in steps of 1$\sigma$). 
As found in \citet[][]{pi09}, the effect
of increasing the threshold is that of decreasing 
the number of extracted clumps.

\begin{figure*}
\centering
\includegraphics[scale=0.5]{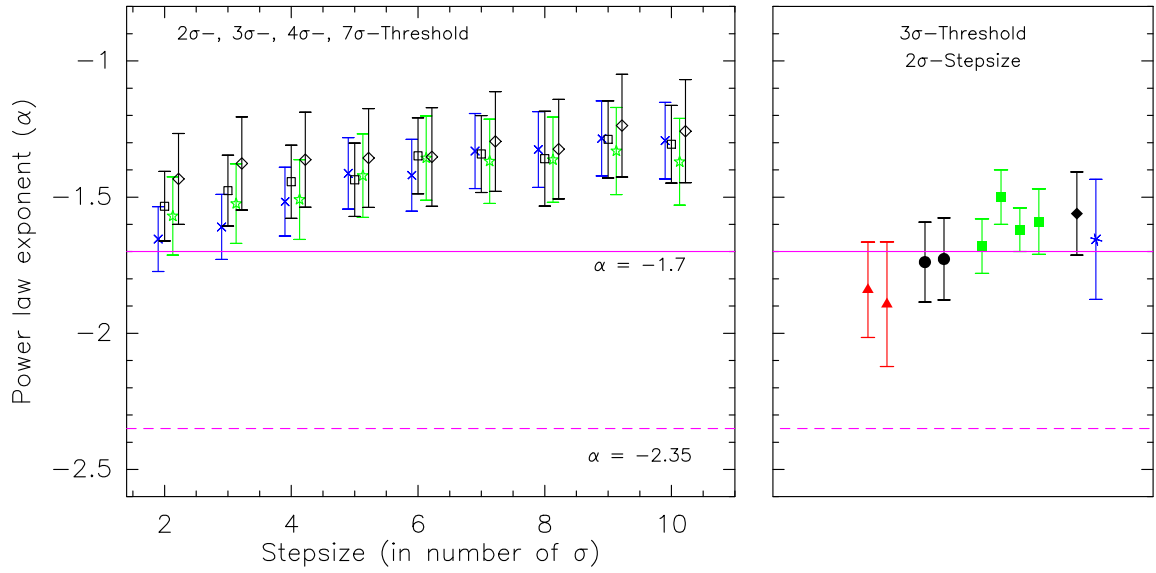}
\caption{Clump mass distribution indices derived from 
{\it Clumpfind} runs for four different thresholds
 (2, 3, 4, and 7$\sigma$) and  nine stepsizes (2--10$\sigma$), assuming one 
temperature of $T=18$~K ({\it left}) and for a given threshold (3$\sigma$) and 
stepsize (2$\sigma$) making different assumptions as mentioned in 
Appendix~\ref{appendix1} ({\it right}). 
 }
\label{Fplotindex}
\end{figure*}

In the {\it left-hand} panel of Fig.~\ref{Fplotindex}, we plot the power-law 
index ($\alpha$) as a function
of stepsize. Same symbols represent runs for a given threshold, i.e.,
crosses, squares, stars, and diamonds represent 2, 3, 4, and 7$\sigma$
thresholds, respectively.
The values of $\alpha$ are estimated for a single temperature ($T=18$~K).
We see that as the threshold and the stepsize increase, the
mass distribution becomes shallower.

Other tests for a given threshold (3$\sigma$) and stepsize (2$\sigma$) are
plotted in the {\it right-hand} panel in Fig.~\ref{Findices}:

\begin{itemize}
\item Red triangles: we removed extended emission from each map by 
using the median filtering technique that calculates the 
median within a box of a given size. We used a box of about ten beams per side.
Median maps were subtracted from the original maps. We then extracted
clumps from these maps and estimated the indices by assuming one temperature 
($T=18$~K) for one case and two temperatures (18~K and 30~K)
for the other. 

\item Black circles: we artificially added 1$\sigma$ rms noise to each map, 
then extracted the clumps, and
used one temperature ($T=18$~K) for one case and two temperatures (18~K and 30~K)
in the other.

\item Green squares: they correspond to the values we present in 
Fig.~\ref{Falldndm-gcl-clf}, which includes two values using the
{\it Gaussclumps} algorithm. 

\item Black diamond: we obtained this value assuming two 
temperatures (18~K and 30~K), and it does not include the clumps that might
lie at very different distances (see Sect.~\ref{comments}). 
The number of clumps lying at different distance would corresponds to less 
than 4\%.

\item Blue asterisk: the same as above but for clumps that do not
have the  signposts of star formation.

\end{itemize}
 
The obtained indices, plotted in the {\it right-hand} panel in 
Fig.~\ref{Findices},
have a mean and standard deviation of $\alpha = -1.68$ and 0.15, respectively.

\end{appendix}

\end{document}